\documentclass[epjST]{svjour}
\pdfoutput=1 
\usepackage{graphics}
\usepackage[dvipsnames]{xcolor}
\usepackage{empheq}
\usepackage{graphicx}
\usepackage{dcolumn}
\usepackage{bm}
\usepackage{hyperref}
\usepackage{lineno}
\usepackage{xspace}
\usepackage{units}
\usepackage{blindtext}
\usepackage[percent]{overpic}

\hypersetup{
    pdfnewwindow=true,      
    colorlinks=true,       
    linkcolor=blue,          
    citecolor=blue,        
    filecolor=blue,      
    urlcolor=blue           
}

\newcommand{\antinue}{$\bar{\nu}_\textrm{e}$\xspace}
\newcommand{\antinumu}{$\bar{\nu}_\mu$\xspace}
\newcommand{\bjx}{Bjorken-$x$\xspace}
\newcommand{\dat}{$\delta\alpha_\textrm{T}$\xspace}

\newcommand{\dpt}{$\delta p_\textrm{T}$\xspace}
\newcommand{\dpty}{$\delta p_\textrm{Ty}$\xspace}
\newcommand{\eav}{$E_\textrm{avail}$\xspace}
\newcommand{\ele}{$\textrm{e}$\xspace}
\newcommand{\enu}{$E_\nu$\xspace}
\newcommand{\gevc}{GeV/$c$\xspace}
\newcommand{\gevcc}{GeV/$c^2$\xspace}

\newcommand{\kplus}{$\textrm{K}^+$\xspace}
\newcommand{\ma}{$M_\textrm{A}$\xspace}
\newcommand{\mevc}{MeV/$c$\xspace}
\newcommand{\neutron}{\textrm{n}}
\newcommand{\nue}{$\nu_\textrm{e}$\xspace}
\newcommand{\nuA}{$\nu$-A\xspace}

\newcommand{\numu}{$\nu_\mu$\xspace}
\newcommand{\nuanu}{\smash{\overset{\scalebox{.3}{(}\raisebox{-1.7pt}{--}\scalebox{.3}{)}}{\nu}}\xspace}
\newcommand{\numuanumu}{\smash{\overset{\scalebox{.3}{(}\raisebox{-1.7pt}{--}\scalebox{.3}{)}}{\nu}}_\mu\xspace}

\newcommand{\pt}{$p_\textrm{T}$\xspace}
\newcommand{\pz}{$p_{||}$\xspace}
\newcommand{\pn}{$p_\textrm{n}$\xspace}
\newcommand{\proton}{\textrm{p}}

\newcommand{\qeqs}{$Q^2_\textrm{QE}$\xspace}
\newcommand{\tabs}{$\left|t\right|$\xspace}
\newcommand{\tpth}{2p2h\xspace}
\newcommand{\vdpt}{$\delta\vec{p}_\textrm{T}$\xspace}

\newcommand{\dune}{DUNE\xspace}
\newcommand{\geant}{\textsc{GEANT4}\xspace}
\newcommand{\genie}{\textsc{GENIE}\xspace}
\newcommand{\gibuu}{\textsc{GiBUU}\xspace}
\newcommand{\minerva}{MINERvA\xspace}
\newcommand{\minos}{MINOS\xspace}
\newcommand{\mnv}{\textsc{MnvGENIE}\xspace}
\newcommand{\neut}{\textsc{NEUT}\xspace}
\newcommand{\nova}{NOvA\xspace}
\newcommand{\nuwro}{\textsc{NuWro}\xspace}
\newcommand{\numi}{NuMI\xspace}

\newcommand{\triplewidth}{\linewidth}
\newcommand{\doublewidth}{0.786\linewidth}
\newcommand{\singlewidth}{0.49\linewidth}
\newcommand{\halfwidth}{0.49\linewidth}

\definecolor{mycol}{rgb}{1.0, 0.49, 0.0}

\begin{document}

\title{Exploring Neutrino-Nucleus Interactions in the GeV Regime using \minerva}

\newcommand{\Rutgers}{Rutgers, The State University of New Jersey, Piscataway, New Jersey 08854, USA}
\newcommand{\Hampton}{Hampton University, Dept. of Physics, Hampton, VA 23668, USA}
\newcommand{\Dortmund}{Institute of Physics, Dortmund University, 44221, Germany }
\newcommand{\Otterbein}{Department of Physics, Otterbein University, 1 South Grove Street, Westerville, OH, 43081 USA}
\newcommand{\JMU}{James Madison University, Harrisonburg, Virginia 22807, USA}
\newcommand{\Florida}{University of Florida, Department of Physics, Gainesville, FL 32611}
\newcommand{\UCIrvine}{Department of Physics and Astronomy, University of California, Irvine, Irvine, California 92697-4575, USA}
\newcommand{\CBPF}{Centro Brasileiro de Pesquisas F\'{i}sicas, Rua Dr. Xavier Sigaud 150, Urca, Rio de Janeiro, Rio de Janeiro, 22290-180, Brazil}
\newcommand{\PUCP}{Secci\'{o}n F\'{i}sica, Departamento de Ciencias, Pontificia Universidad Cat\'{o}lica del Per\'{u}, Apartado 1761, Lima, Per\'{u}}
\newcommand{\INRM}{Institute for Nuclear Research of the Russian Academy of Sciences, 117312 Moscow, Russia}
\newcommand{\Jlab}{Jefferson Lab, 12000 Jefferson Avenue, Newport News, VA 23606, USA}
\newcommand{\Pittsburgh}{Department of Physics and Astronomy, University of Pittsburgh, Pittsburgh, Pennsylvania 15260, USA}
\newcommand{\Guanajuato}{Campus Le\'{o}n y Campus Guanajuato, Universidad de Guanajuato, Lascurain de Retana No. 5, Colonia Centro, Guanajuato 36000, Guanajuato M\'{e}xico.}
\newcommand{\Athens}{Department of Physics, University of Athens, GR-15771 Athens, Greece}
\newcommand{\Tufts}{Physics Department, Tufts University, Medford, Massachusetts 02155, USA}
\newcommand{\WM}{Department of Physics, William \& Mary, Williamsburg, Virginia 23187, USA}
\newcommand{\FNAL}{Fermi National Accelerator Laboratory, Batavia, Illinois 60510, USA}
\newcommand{\Purdue}{Department of Chemistry and Physics, Purdue University Calumet, Hammond, Indiana 46323, USA}
\newcommand{\MCLA}{Massachusetts College of Liberal Arts, 375 Church Street, North Adams, MA 01247}
\newcommand{\UMD}{Department of Physics, University of Minnesota -- Duluth, Duluth, Minnesota 55812, USA}
\newcommand{\Northwestern}{Northwestern University, Evanston, Illinois 60208}
\newcommand{\UNI}{Facultad de Ciencias, Universidad Nacional de Ingenier\'{i}a, Apartado 31139, Lima, Per\'{u}}
\newcommand{\Rochester}{University of Rochester, Rochester, New York 14627 USA}
\newcommand{\Austin}{Department of Physics, University of Texas, 1 University Station, Austin, Texas 78712, USA}
\newcommand{\USM}{Departamento de F\'{i}sica, Universidad T\'{e}cnica Federico Santa Mar\'{i}a, Avenida Espa\~{n}a 1680 Casilla 110-V, Valpara\'{i}so, Chile}
\newcommand{\Geneva}{University of Geneva, 1211 Geneva 4, Switzerland}
\newcommand{\Chicago}{Enrico Fermi Institute, University of Chicago, Chicago, IL 60637 USA}
\newcommand{\hired}{}
\newcommand{\OregonState}{Department of Physics, Oregon State University, Corvallis, Oregon 97331, USA}
\newcommand{\oxford}{University of Oxford, Department of Physics, Oxford, OX1 3PJ United Kingdom}
\newcommand{\umiss}{University of Mississippi, Oxford, Mississippi 38677, USA}
\newcommand{\upenn}{Department of Physics and Astronomy, University of Pennsylvania, Philadelphia, PA 19104}
\newcommand{\AMU}{AMU Campus, Aligarh, Uttar Pradesh 202001, India}
\newcommand{\wroclaw}{University of Wroclaw, plac Uniwersytecki 1, 50-137 Wroa\l{}aw, Poland}
\newcommand{\Mohali}{Department of Physical Sciences, IISER Mohali, Knowledge City, SAS Nagar, Mohali - 140306, Punjab, India}
\newcommand{\CINVESTAV}{Departamento de Fisica Col. San Pedro Zacatenco, 07360 Mexico, DF, Av. Instituto PolitÃ©cnico Nacional, Mexico}
\newcommand{\york}{York University, Department of Physics and Astronomy, Toronto, Ontario, M3J 1P3 Canada}
\newcommand{\ND}{Department of Physics, University of Notre Dame, Notre Dame, Indiana 46556, USA}
\newcommand{\ICL}{The Blackett Laboratory,  Imperial College London,  London SW7 2BW, United Kingdom}

\newcommand{\mateusfcarneiroThanks}{Now at Brookhaven National Laboratory}
\newcommand{\finerThanks}{Now at Los Alamos National Laboratory}
\newcommand{\emilymaherThanks}{Department of Physics}
\newcommand{\bamThanks}{Now at University of Minnesota}

\newcommand{\idoxford}{1}
\newcommand{\idWM}{2}
\newcommand{\idAMU}{3}
\newcommand{\idGuanajuato}{4}
\newcommand{\idPUCP}{5}
\newcommand{\idOregonState}{6}
\newcommand{\idFNAL}{7}
\newcommand{\idRochester}{8}
\newcommand{\idCBPF}{9}
\newcommand{\idTufts}{10}
\newcommand{\idUMD}{11}
\newcommand{\idyork}{12}
\newcommand{\idMohali}{13}
\newcommand{\idICL}{14}
\newcommand{\idupenn}{15}
\newcommand{\idMCLA}{16}
\newcommand{\idPittsburgh}{17}
\newcommand{\idUSM}{18}
\newcommand{\idFlorida}{19}
\newcommand{\idRutgers}{20}
\newcommand{\idumiss}{21}
\newcommand{\idUNI}{22}

\author{\begin{center}
X.-G.~Lu                         \inst{\idoxford}\fnmsep\thanks{\email{xianguo.lu@physics.ox.ac.uk}} \and
Z.~~Ahmad~Dar                    \inst{\idWM,\idAMU}\and
F.~Akbar                         \inst{\idAMU}\and
D.A.~Andrade                     \inst{\idGuanajuato}\and
M.~V.~Ascencio                   \inst{\idPUCP}\and
G.D.~Barr                        \inst{\idoxford}\and
A.~Bashyal                       \inst{\idOregonState}\and
L.~Bellantoni                    \inst{\idFNAL}\and
A.~Bercellie                     \inst{\idRochester}\and
M.~Betancourt                    \inst{\idFNAL}\and
A.~Bodek                         \inst{\idRochester}\and
J.~L.~Bonilla                    \inst{\idGuanajuato}\and
H.~Budd                          \inst{\idRochester}\and
G.~Caceres                       \inst{\idCBPF}\and
T.~Cai                           \inst{\idRochester}\and
M.F.~Carneiro\inst{\idOregonState,\idCBPF}\fnmsep\thanks{\mateusfcarneiroThanks}  \and
H.~da~Motta                      \inst{\idCBPF}\and
G.A.~D\'{i}az                    \inst{\idRochester}\and
J.~Felix                         \inst{\idGuanajuato}\and
L.~Fields                        \inst{\idFNAL}\and
A.~Filkins                       \inst{\idWM}\and
R.~Fine\inst{\idRochester}\fnmsep\thanks{\finerThanks}     \and
A.M.~Gago                        \inst{\idPUCP}\and
H.~Gallagher                     \inst{\idTufts}\and
S.M.~Gilligan                    \inst{\idOregonState}\and
R.~Gran                          \inst{\idUMD}\and
D.A.~Harris                      \inst{\idyork,\idFNAL}\and
S.~Henry                         \inst{\idRochester}\and
D.~Jena                          \inst{\idFNAL}\and
S.~Jena                          \inst{\idMohali}\and
J.~Kleykamp                      \inst{\idRochester}\and
A.~Klustová                      \inst{\idICL}\and
M.~Kordosky                      \inst{\idWM}\and
D.~Last                          \inst{\idupenn}\and
A.~Lozano                        \inst{\idCBPF}\and
E.~Maher\inst{\idMCLA}\fnmsep\thanks{\emilymaherThanks}  \and
S.~Manly                         \inst{\idRochester}\and
W.A.~Mann                        \inst{\idTufts}\and
C.~Mauger                        \inst{\idupenn}\and
K.S.~McFarland                   \inst{\idRochester}\and
A.M.~McGowan                     \inst{\idRochester}\and
B.~Messerly\inst{\idPittsburgh}\fnmsep\thanks{\bamThanks}   \and
J.~Miller                        \inst{\idUSM}\and
J.G.~Morf\'{i}n                  \inst{\idFNAL}\and
D.~Naples                        \inst{\idPittsburgh}\and
J.K.~Nelson                      \inst{\idWM}\and
C.~Nguyen                        \inst{\idFlorida}\and
A.~Olivier                       \inst{\idRochester}\and
V.~Paolone                       \inst{\idPittsburgh}\and
G.N.~Perdue                      \inst{\idFNAL,\idRochester}\and
K.-J.~Plows                      \inst{\idoxford}\and
M.A.~Ram\'{i}rez                 \inst{\idupenn,\idGuanajuato}\and
R.D.~Ransome                     \inst{\idRutgers}\and
H.~Ray                           \inst{\idFlorida}\and
P.A.~Rodrigues                   \inst{\idRochester,\idumiss,\idoxford}  
D.~Ruterbories                   \inst{\idRochester}\and
H.~Schellman                     \inst{\idOregonState}\and
C.J.~Solano~Salinas              \inst{\idUNI}\and
H.~Su                            \inst{\idPittsburgh}\and
M.~Sultana                       \inst{\idRochester}\and
V.S.~Syrotenko                   \inst{\idTufts}\and
E.~Valencia                      \inst{\idWM,\idGuanajuato}\and
A.V.~Waldron                     \inst{\idICL}\and
D.~Wark                          \inst{\idoxford}\and
A.~Weber                         \inst{\idoxford}\and
K.~Yang                          \inst{\idoxford}\and
L.~Zazueta                       \inst{\idWM}\\
(The MINERvA Collaboration)
\end{center}
}

\institute{
\oxford\and
\WM\and
\AMU\and
\Guanajuato\and
\PUCP\and
\OregonState\and
\FNAL\and
\Rochester\and
\CBPF\and
\Tufts\and
\UMD\and
\york\and
\Mohali\and
\ICL\and
\upenn\and
\MCLA\and
\Pittsburgh\and
\USM\and
\Florida\and
\Rutgers\and
\umiss\and
\UNI
}

\abstract{
With the advance of particle accelerator and detector technologies, the neutrino physics landscape is rapidly expanding. As neutrino oscillation experiments enter the intensity and precision frontiers, neutrino-nucleus interaction measurements are providing crucial input. \minerva is an experiment at Fermilab dedicated to the study of neutrino-nucleus interactions in the regime
of incident neutrino energies from one to few GeV.
The experiment recorded neutrino and antineutrino scattering data with the \numi beamline from 2009 to 2019 using the Low-Energy and Medium-Energy beams that peak at  \unit[3]{GeV} and \unit[6]{GeV}, respectively.  This article reviews the 
broad spectrum of interesting nuclear and particle physics that \minerva investigations have illuminated.
The newfound, detailed knowledge of neutrino interactions with nuclear targets thereby obtained is proving essential to continued progress in the neutrino physics sector.
}  

\maketitle

\setcounter{tocdepth}{2}
\tableofcontents

\section{Introduction} \label{intro}
 
Neutrinos with energies of a few GeV are involved in many different ways among phenomena that
present opportunities to probe fundamental aspects of physical reality.  Neutrinos produced in accelerators play a central role in precision measurements of the oscillation parameters such as the Dirac CP-violating phase that may be present in the neutrino flavor mixing matrix~\cite{Abe:2019vii,Acero:2019ksn,Abi:2020wmh,Abe:2018uyc}. Measurement of a non-zero Dirac phase could unlock the mystery of the matter-antimatter asymmetry of the Universe. Neutrino beams serve as potential sources of beyond-Standard-Model (BSM) particles, such as light dark matter and heavy neutral leptons~\cite{Abe:2019kgx,Abi:2020kei}.  On the other hand, neutrinos could impede the discovery of such new forms of matter by mimicking their BSM signatures. This possibility exists
because some neutrino SM processes in detector materials have aspects that are poorly known. 
Atmospheric neutrinos that are born in cosmic-ray-induced hadronic cascades in the upper atmosphere propagate through the Earth~\cite{Abe:2017aap}, presenting complications as well as opportunities
for new physics searches.   Atmospheric neutrinos oscillate, and their oscillations undergo highly
interesting alterations due to propagation through a matter field.   However, 
these highly penetrating particles also create background to rare-event searches (such as proton decay~\cite{Abe:2014mwa,An:2015jdp}) in deep underground experiments. 
Understanding how a neutrino interacts with a nucleus is essential for exploiting these opportunities. Given that current GeV-neutrino sources  
(accelerators or atmospheric) are not monoenergetic, these energy-sensitive interactions are convolved with the neutrino flux, causing major systematic uncertainties in precision measurements. 

Neutrino-nucleus (\nuA) interactions arise not only from the primary nucleon-level interaction, but also from the effect that the nuclear environment exerts on the initial-state nucleons and the final-state particles. 
Since a theory of the complete nuclear response in neutrino-nucleus interactions in the few-GeV regime of incident neutrino energy is yet to be developed~\cite{Alvarez-Ruso:2017oui}, comprehensive \nuA measurements are needed 
to guide and benchmark the development of models.
\minerva (Main INjector ExpeRiment for \nuA) at Fermilab is a dedicated experiment to illuminate the interplay between hadronic and nuclear degrees of freedom in \nuA interactions and to measure aspects of intranuclear dynamics that are prerequisites for precision neutrino oscillation measurements.

\minerva received the \numi (Neutrinos-from-the-Main-Injector) beam at a distance of \unit[1]{km} from the target of the 120-GeV primary proton beam at Fermilab. In the Low-Energy (LE) beam configurations operated between 2009 and 2012,  both the \numu and \antinumu fluxes peak at $\sim$\unit[3]{GeV}, while in the Medium-Energy (ME) configurations used between 2013 and 2019, the fluxes peak at $\sim$\unit[6]{GeV}.  In both beams, there is a high-energy component that extends beyond \unit[50]{GeV}. The data collected by \minerva correspond to 4.0 (1.7) and 12.1 (12.4) times $10^{20}$ protons on target (POT) for the LE and ME \numu (\antinumu) configurations, respectively. 
In the Sections below, the neutrino interaction physics investigated by \minerva is reviewed,
with the main focus being the techniques developed and measurements reported that are based
on the Low-Energy data set. 
 
\section{\minerva Experiment and Flux Predictions}

The \minerva detector~\cite{Aliaga:2013uqz} utilizes extruded plastic scintillator as its tracking medium.   Most of the active mass is located in its central polystyrene target whose 5.4-t fiducial volume serves as a charged-particle tracker. The upstream section of the detector consists of a series of passive targets (helium, carbon, water, iron, and lead) interleaved with tracking planes. Sampling calorimeters surround both passive and active target regions. 
Muons produced by \nuA charged-current (CC) interactions in the tracker or in the upstream
targets exit the downstream end of the tracker. These muons may then enter and propagate through the magnetized \minos near detector~\cite{MINOS:2008hdf} located \unit[2]{m} downstream, allowing their trajectories to be momentum-analyzed.

The downstream tracking of muons in the \minos near detector provides an \unit[8]{\%} muon momentum resolution (at \unit[5]{\gevc}).  This complements the three-dimensional tracking and energy-loss measurement of final-state particles that is afforded by the \minerva tracker. In the tracker, 
the momentum resolution for protons is \unit[2]{\%} at \unit[1]{\gevc} with a 450-\mevc tracking threshold. The hadronic energy response of the detector was calibrated using test-beam measurements~\cite{Aliaga:2015aqe}, with pion calorimetric energy resolution in the range \unit[20-30]{\%}.  Furthermore, the tracker is large compared to the  interaction length of the neutrons (approximately \unit[10]{cm} at \unit[20]{MeV}) produced in \nuA interactions. The 1.5-MeV detection threshold for measuring energy deposit allows interacting neutrons to be registered for a time-of-flight measurement~\cite{Elkins:2019vmy}. The neutron timing resolution is \unit[4.5]{ns} where the hit resolution alone is \unit[3]{ns} from electronics effects.

\begin{figure}[!ht]
\begin{center}
\begin{overpic}[trim={0.2cm 0 0cm 1.15cm},clip,width=\singlewidth]{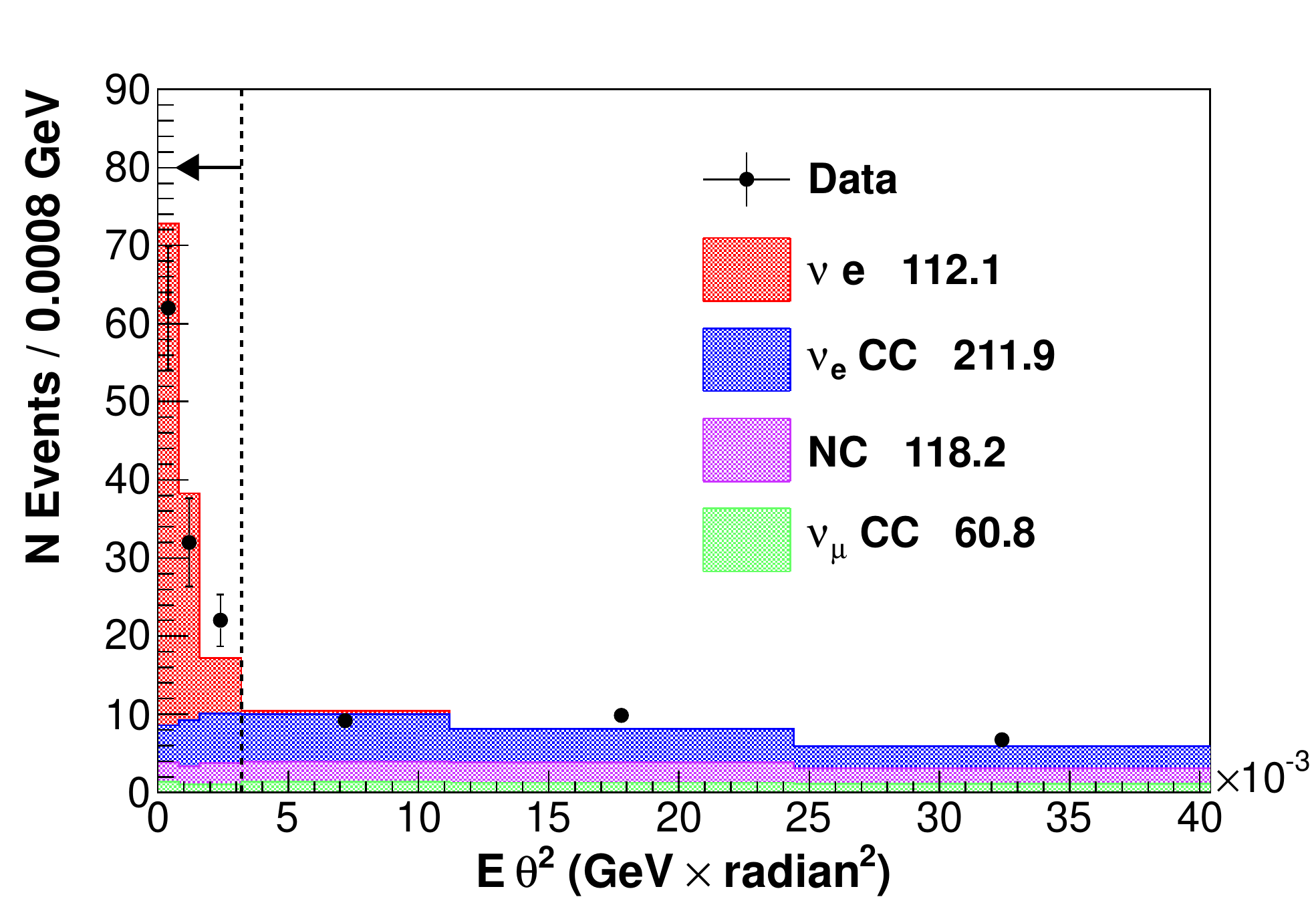}
\put (25,35) {(a) LE}
\end{overpic}
\begin{overpic}[trim={0.2cm 0 1.5cm 1.1cm},clip,width=\singlewidth]{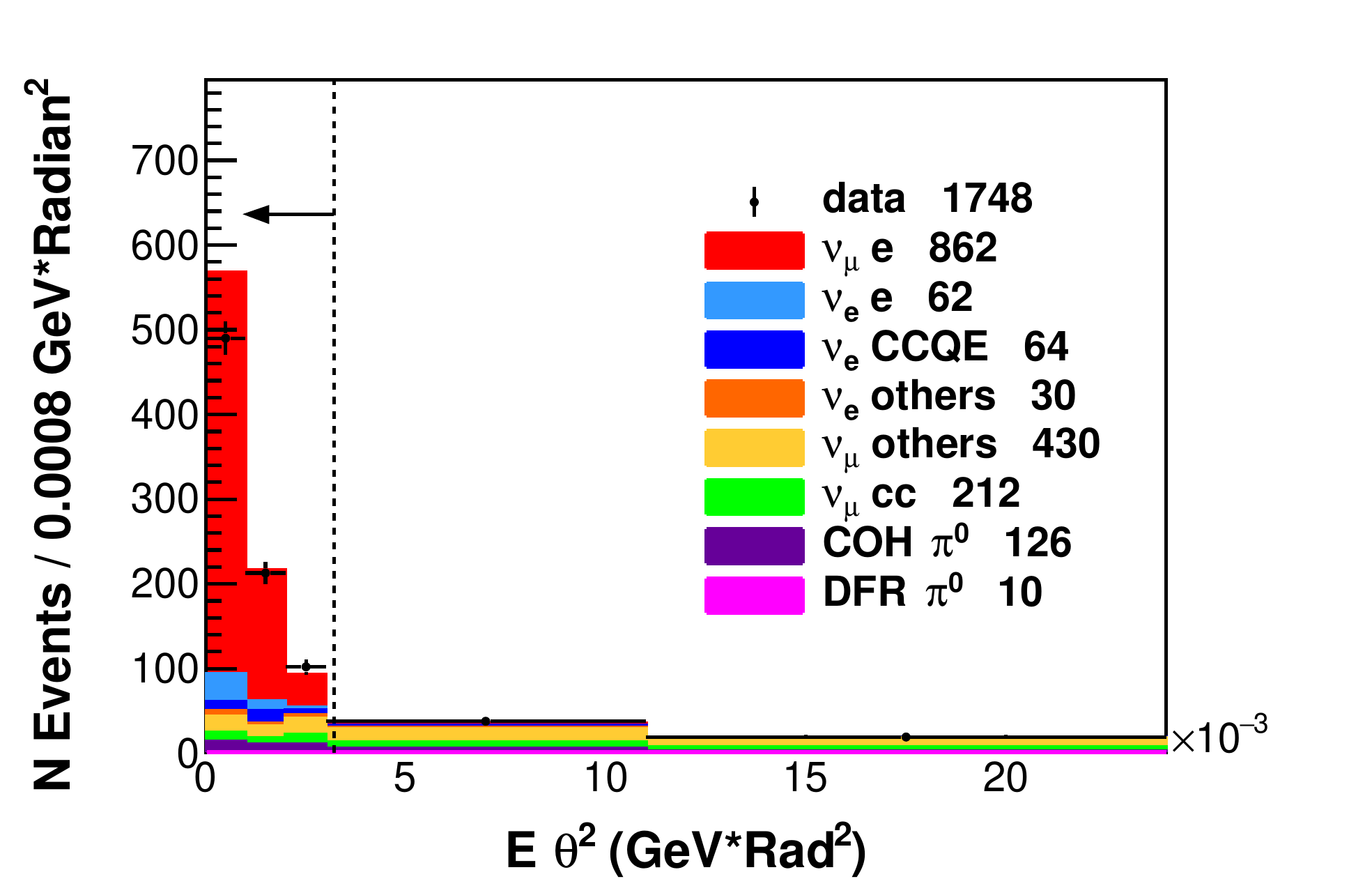}
\put (30,35) {(b) ME}
\end{overpic}
\caption{Reconstructed electron energy ($E$) times the square of the electron angle with respect to the beam ($\theta$) for the neutrino-electron elastic scattering candidates with the \numi (a) LE and (b) ME configurations. Left of the dash lines are the selected samples.  Stacked histograms are the simulated event contributions using the \textit{a priori} flux models (NC refers to other neutral-current events that are background. COH and DFR stand for coherent and diffractive production, respectively). Figures from Refs.~\cite{Park:2015eqa,Valencia:2019mkf}.}
\label{fig:nuelas}
\end{center}
\end{figure}

The \numi beam fluxes used by \minerva are modeled with \geant predictions that are adjusted to match hadron production data~\cite{Aliaga:2016oaz}. The large fiducial volume of the \minerva tracker and the intense beam fluxes make it possible to use neutrino scattering on atomic electrons, $\numuanumu\ele^{-}\rightarrow\numuanumu\ele^{-}$, to further constrain the flux predictions. 
In the data from the LE (ME) NuMI beam configurations,
135 (810) neutrino-electron scattering events were identified~\cite{Park:2015eqa,Valencia:2019mkf}.
The constraints thereby provided reduce the \numu flux normalization in the \textit{a priori} prediction by \unit[6]{\%} for LE, and  \unit[10]{\%} for ME (Fig.~\ref{fig:nuelas}).  Moreover, the  
uncertainty at the flux peak is reduced from 9 to \unit[6]{\%} for LE, and from  8 to \unit[4]{\%} for ME. 
In addition, the ``low-$\nu$" method has been used to constrain the flux shape~\cite{DeVan:2016rkm,Ren:2017xov}.  The latter method exploits the minimal neutrino energy dependence of the inclusive charged-current cross section at low hadronic recoil energy. 

\section{Incoherent Neutrino-Nucleus Interactions}\label{intranu}

In the few-GeV regime, neutrino-nucleus cross sections are dominated by incoherent processes in which the constituent nucleons can be ejected, possibly accompanied by pions and other mesons. These primary processes are quasielastic (QE), baryon resonance  production (RES) including non-resonant background, and deep inelastic scattering (DIS), in ascending order of excitation by incident 
neutrinos of increasing lab-frame energies.
The initial energy and momentum of the struck nucleon and its correlation with other nucleons (including long-range correlations and two-particle-two-hole excitations, or \tpth~\cite{Marteau:1999kt,Martini:2010ex,Nieves:2011pp,Gran:2013kda,Benhar:2015ula}) contribute to the initial-state conditions.  The primary hadronic final-state particles propagate inside the remnant nucleus and may participate in final-state intranuclear interactions (FSI).
The latter interactions may further excite the remaining system, causing nucleon emission or even spallation. 
Because pions can be absorbed or created during FSI, there is no unique experimental signature for a given primary process. As a result, the particle content and energy budget of a \nuA interaction varies with the initial and final states together with the primary reaction. 

By restricting the final-state topology, \minerva can examine exclusive and semi-inclusive
reactions such as mesonless (i.e., quasielastic-like), pion, and kaon production; the
experiment can study inclusive scattering as well.    The exclusive-channel studies must
take into account pion absorption through FSI, which enables resonance production and DIS reactions to be present in quasielastic-like topologies.  

The \minerva data enable the elucidation of one-particle-one-hole mechanisms that are generally used in \nuA scattering models, and they also allow examination of the 
significant \tpth contributions to the quasielastic-like process.  
Concerning the latter, \minerva has identified a \tpth-like enhancement that   
simultaneously describes both neutrino and antineutrino scattering data at the kinematic region between quasielastic and resonance production. These unmodeled additional event rates are likely an admixture of all three reactions and their precise nature is still under study.  

Decoupling the primary process and medium effects is challenging, especially in a wide-band neutrino beam where the  
neutrino energy (\enu) is unknown. 
In first order, reaction \enu-dependence comes from the primary interactions and the final-state momenta and angles depend strongly on \enu. The nuclear response, on the other hand, affects these  elementary distributions as a perturbation; it depends on \enu through the medium coupling to the primary initial-and final-state hadrons. 
With a neutrino beam where the neutrino direction is precisely known, the kinematics projected onto the transverse plane to the neutrino     
will have less dependence on \enu. 

In a charged-current measurement, certain final-state correlations of the lepton and the hadronic system, such as the transverse kinematic imbalance (TKI)~\cite{Lu:2015hea,Lu:2015tcr}, avoid or cancel the primary-level dependence on, for example, \enu and axial form factors, and are directly sensitive to the nuclear response with minimal dependence on the neutrino energy. 
Because exclusivity can be achieved in the transverse plane to the neutrino direction, TKI can probe the hidden dynamics inside the target nucleus.  
\minerva has  systematically explored the potential of TKI to identify the medium properties and interaction dynamics in exclusive processes.

A review of \minerva's measurements of incoherent interactions is presented in this section.
These measurements encompass both exclusive and elastic reactions to inclusive and inelastic processes. 

\subsection{First Measurements of Quasielastic and Quasielastic-Like Cross Sections}\label{sec:1stqe}

The charged-current quasielastic interaction (CCQE) is an important channel in neutrino oscillation experiments. This interaction gives rise to the majority of events in T2K~\cite{Abe:2019vii} and to a sizable
fraction of the events recorded by \nova~\cite{Acero:2019ksn}.  In quasielastic scattering, the neutrino energy can in principle be inferred from the outgoing lepton kinematics.  However, 
this relatively simple interaction incurs significant effects from the nuclear environment.

\begin{figure}[t]
\begin{center}
\begin{overpic}[trim={0.5cm 0 2.9cm 0},clip,width=\halfwidth]{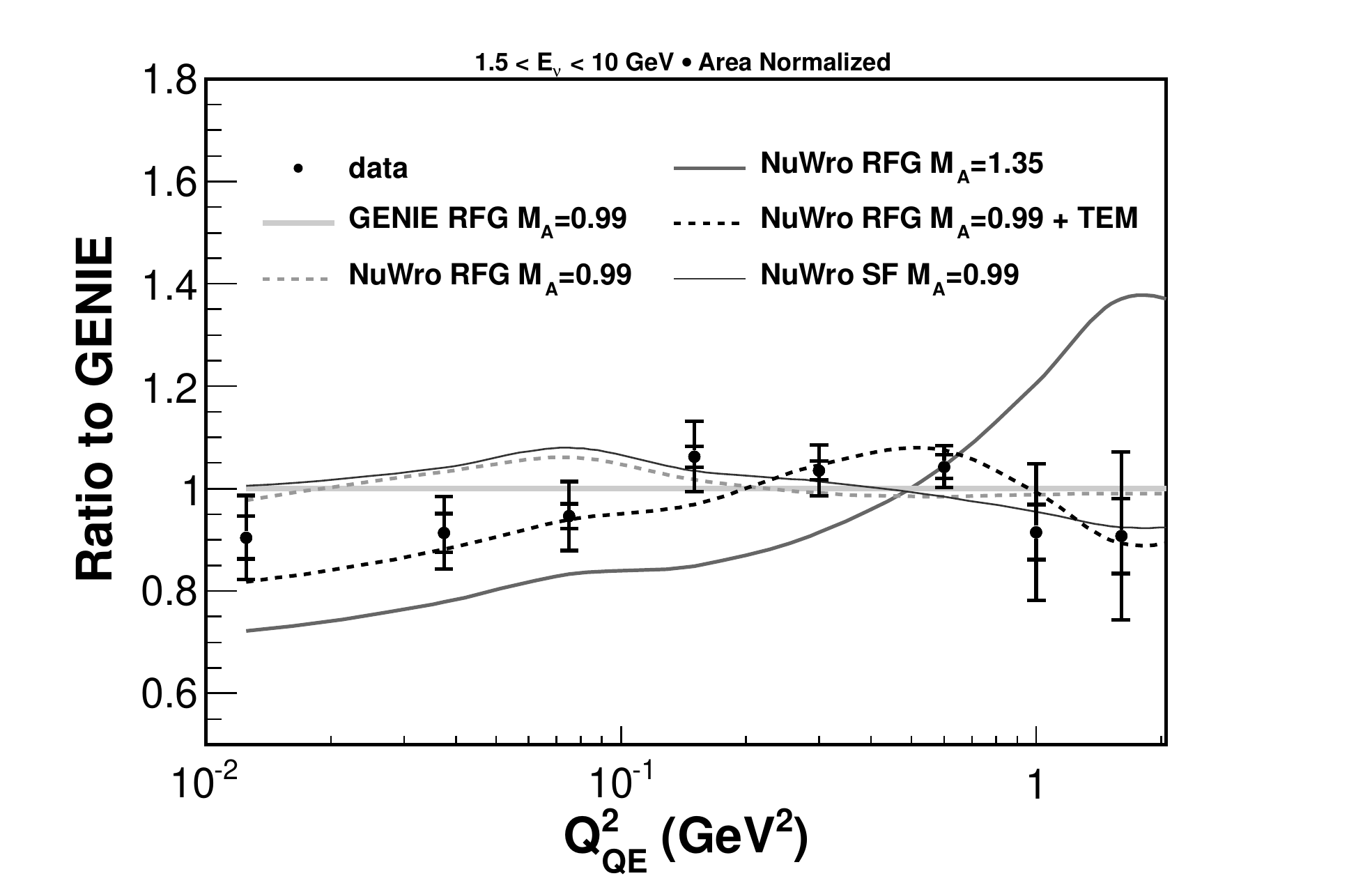}
 \put (20,66) {(a) \numu$\neutron\rightarrow\mu^-\,\proton$}
\end{overpic}
\begin{overpic}[trim={0.5cm 0 2.9cm 0},clip,width=\halfwidth]{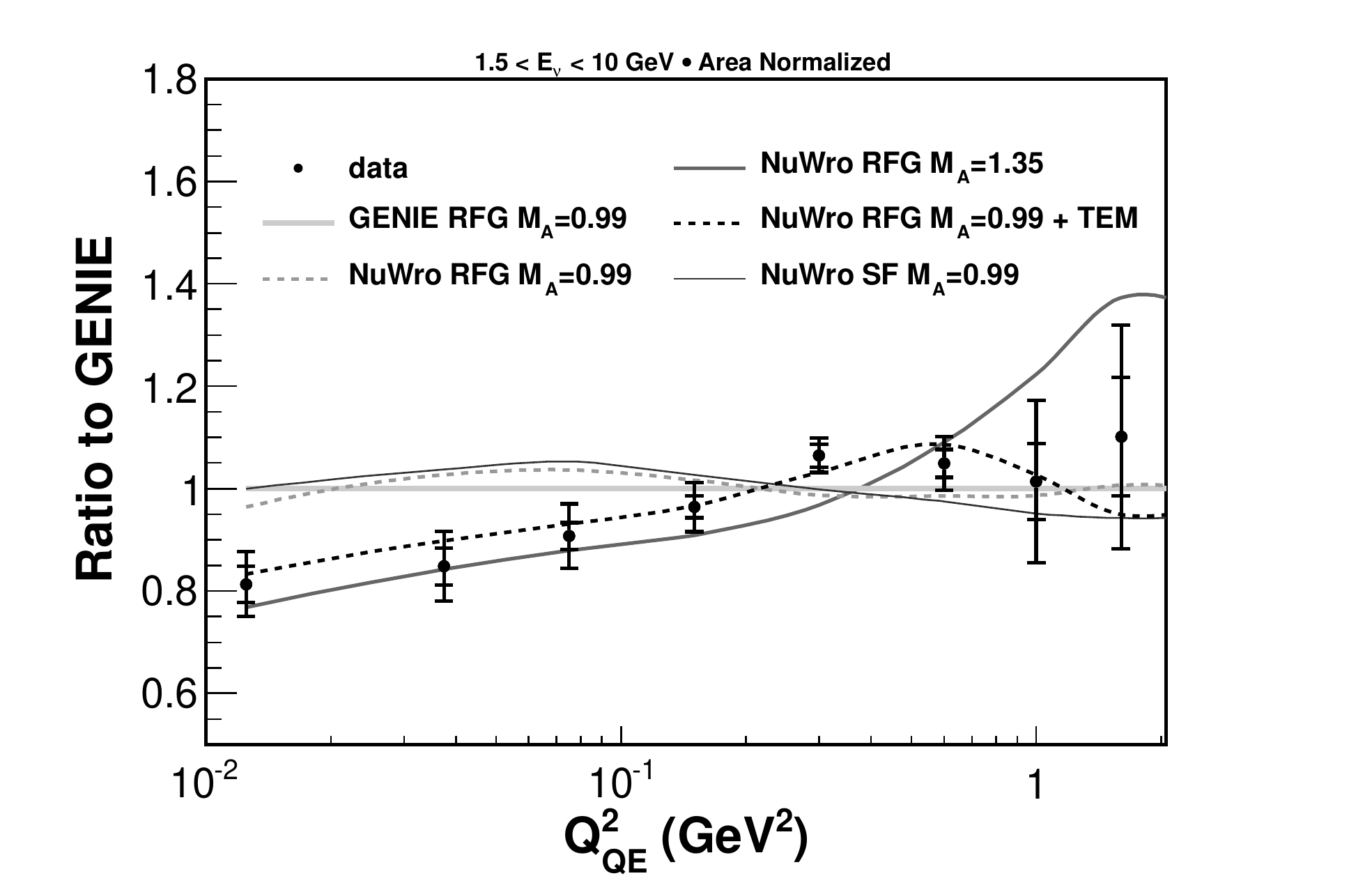}
 \put (20,66) {(b) \antinumu$\proton\rightarrow\mu^+\,\neutron$}
\end{overpic}
\caption{(a) The \numu and (b) \antinumu CCQE cross section in \qeqs. The subscript QE refers to the quasielastic hypothesis that uses only the muon kinematics in the calculation where the target nucleon is assumed to be at rest. The data and model predictions are area-normalized (shape comparison only) and shown as ratios relative to \genie. Figures from Refs.~\cite{Fields:2013zhk,Fiorentini:2013ezn}.}
\label{fig:qeq2}
\end{center}
\end{figure}

Candidate CCQE interactions of neutrinos and antineutrinos on nucleons, \numu$\neutron\rightarrow\mu^-\,\proton$ and  \antinumu$\proton\rightarrow\mu^+\,\neutron$, respectively, were extracted from the Low-Energy data in the first two cross-section measurements 
reported by \minerva~\cite{Fields:2013zhk,Fiorentini:2013ezn}. After subtracting the non-QE background processes, namely resonance production and DIS, the flux-integrated differential cross sections in four-momentum transfer squared, $Q^2$, were compared to model predictions by \genie~\cite{Andreopoulos:2009rq} and \nuwro~\cite{Golan:2012wx} (Fig.~\ref{fig:qeq2}). In the 
neutrino generator predictions, the shape of the $Q^2$ distribution is parameterized by the
axial vector mass, \ma, plus model representations of the nuclear state (relativistic Fermi gas, or RFG~\cite{Bodek:1980ar}, and Spectral Function, or SF~\cite{Benhar:1994hw}). 
Both the neutrino and antineutrino data sets were found to
favor a parametric enhancement in the magnetic form factor (Transverse Enhancement Model, or TEM~\cite{Bodek:2011ps}), in addition to the \emph{at-that-time}-standard choice of RFG and the world-average \ma value of \unit[0.99]{\gevcc}. Becasue TEM is extracted from a fit to electron scattering data to describe the contributions from two-nucleon knock-out processes, these results suggest possible contributions from \tpth in neutrino and antineutrino scattering. This interpretation is further supported by the observed pattern of energy deposits near the interaction vertices (vertex energy)  in both measurements.  

\begin{figure}[t]
\begin{center}
\includegraphics[trim={0 0 2.9cm 1cm},clip, width=\singlewidth]{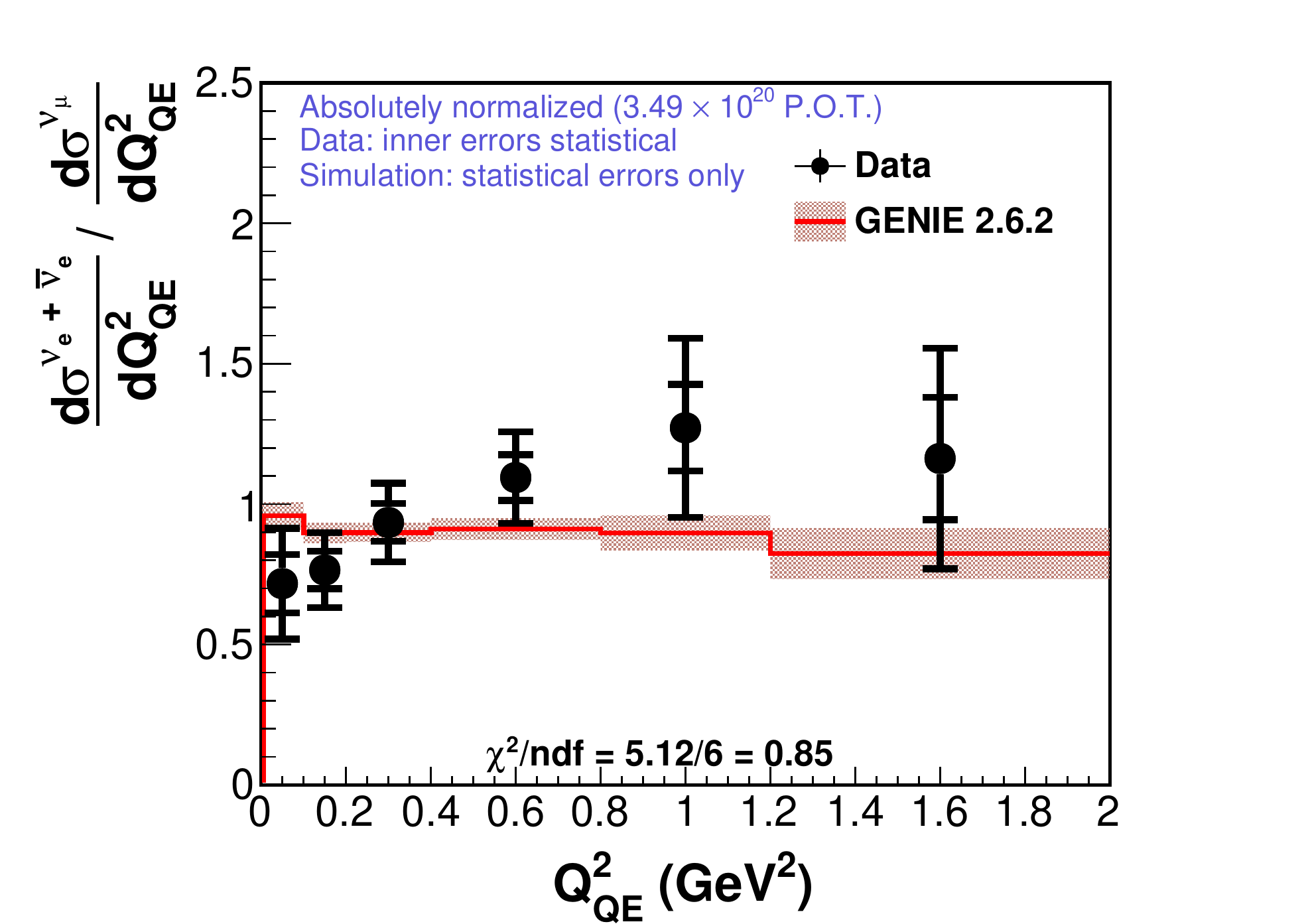}
\caption{Flux-integrated \nue (dominant) and \antinue combined CCQE cross section divided by the \numu counterpart, compared to the \genie prediction. Figure from Ref.~\cite{Wolcott:2015hda}.}
\label{fig:nueqe}
\end{center}
\end{figure}

\minerva also measured the \nue CCQE cross section in the Low-Energy neutrino data set~\cite{Wolcott:2015hda}. In the Standard Model, lepton couplings are universal. However, because the final-state lepton mass is different in \nue versus \numu CC events, the  nuclei respond to slightly different phase spaces---a difference that cannot be ignored in oscillation experiments~\cite{Day:2012gb}. As the CC-induced electrons are reconstructed in the tracker which is charge-insensitive, a flux-averaged cross section including limited contributions from \antinue is obtained. By comparing to the \numu counterpart~\cite{Fiorentini:2013ezn}, it has been shown that the two CCQE cross sections are consistent (Fig.~\ref{fig:nueqe}). Moreover, \genie describes the \nue-to-\numu cross-section ratio within the experimental error which is of order 10-20\%. 

\begin{figure}[!ht]
\begin{center}
\includegraphics[trim={0.8cm 0 2.8cm 1.1cm},clip, width=\singlewidth]{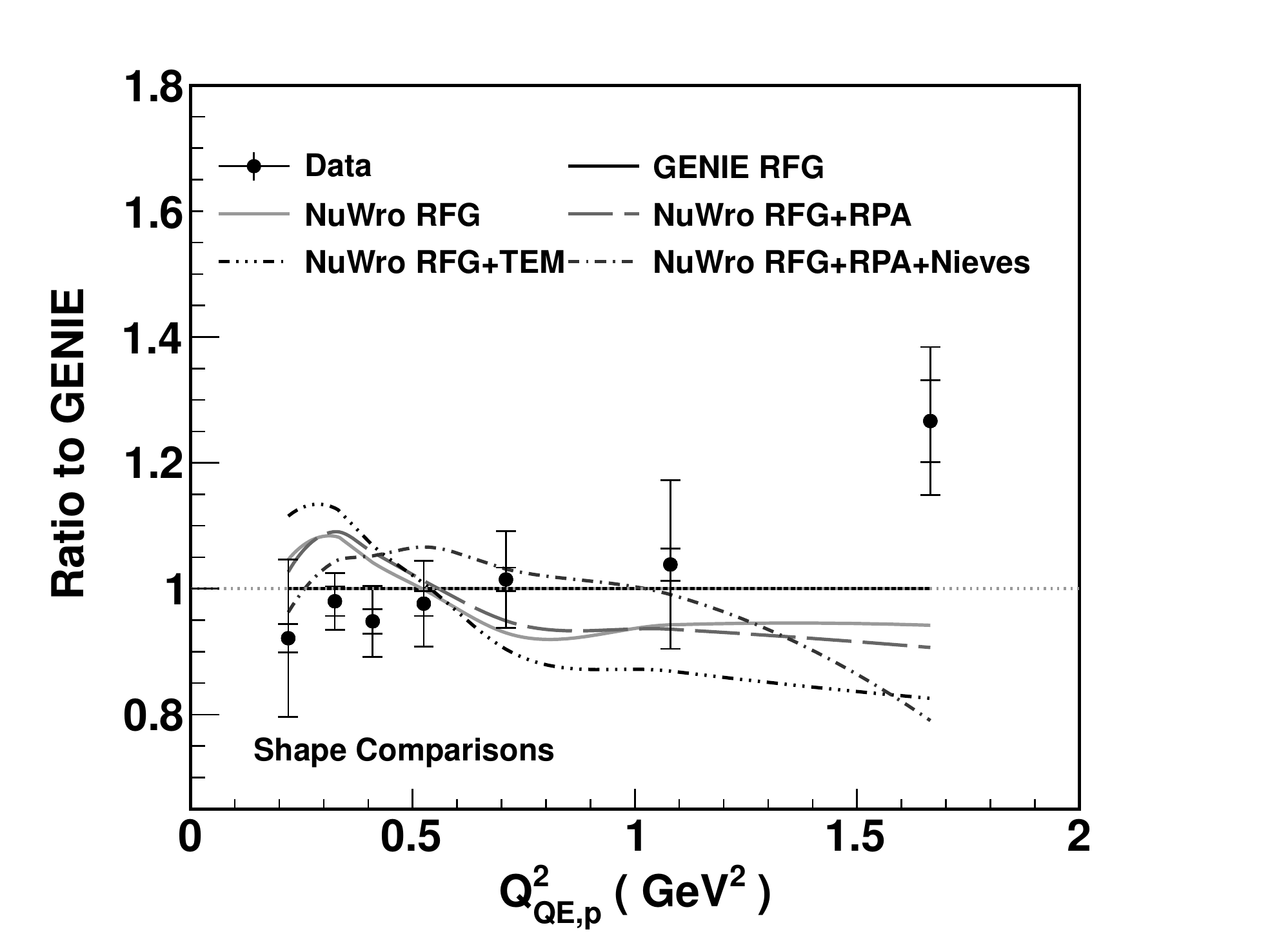}
\caption{The \numu CCQE-like cross section in $Q^2_\textrm{QE,\proton}$. The additional subscript p denotes the pure hadronic calculation only using the final-state proton momentum.
Comparisons are made to generator predictions that include
TEM, Random Phase Approximation (RPA)~\cite{Nieves:2004wx,Graczyk:2003ru}, and the Valencia \tpth model by Nieves~\textit{et al.}~\cite{Nieves:2011yp,Sobczyk:2012ms,Gran:2013kda,Schwehr:2016pvn} (Sect.~\ref{sec:enhance}). Figure from Ref.~\cite{Walton:2014esl}.}
\label{fig:qeq2wproton}
\end{center}
\end{figure}

In \minerva quasielastic measurements, the inelastic background arising from primary pion production that is followed by absorption during FSI must be subtracted. In subsequent \minerva studies of the quasielastic dynamics, the inelastic pionless events are instead considered as part of the 
``QE-like" signal definition in order to mitigate the model uncertainties for pion absorption. 
In \minerva's first QE-like cross-section measurement~\cite{Walton:2014esl}, the $Q^2$ distribution is extracted. In addition to the change of signal definition, the $Q^2$ calculation is completely hadron-based, using only the momentum of the leading final-state proton above the tracking threshold. This was the first \minerva measurement to use reconstructed proton tracks. The distribution is compared to model predictions in order to test different hypotheses concerning initial-state correlations and the data are found to disfavor the modeled correlations (Fig.~\ref{fig:qeq2wproton}). 

\begin{figure}[!ht]
\begin{center}
\begin{overpic}[trim={1.05cm 0.5cm 3cm 1.1cm},clip,width=\halfwidth]{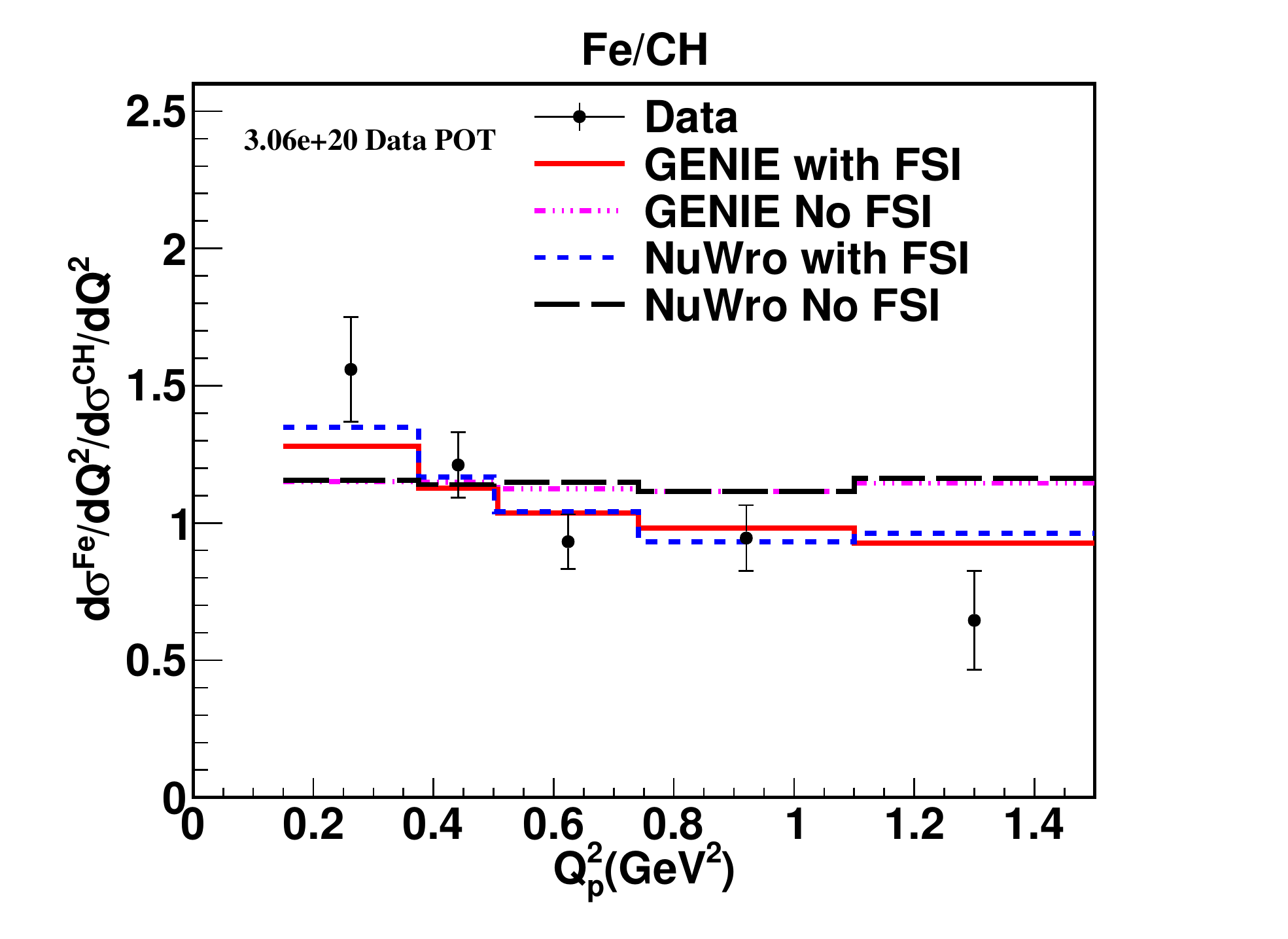}
\put (20,20) {(a) Fe/CH}
\end{overpic}
\begin{overpic}[trim={1.05cm 0.5cm 3cm 1.1cm},clip,width=\halfwidth]{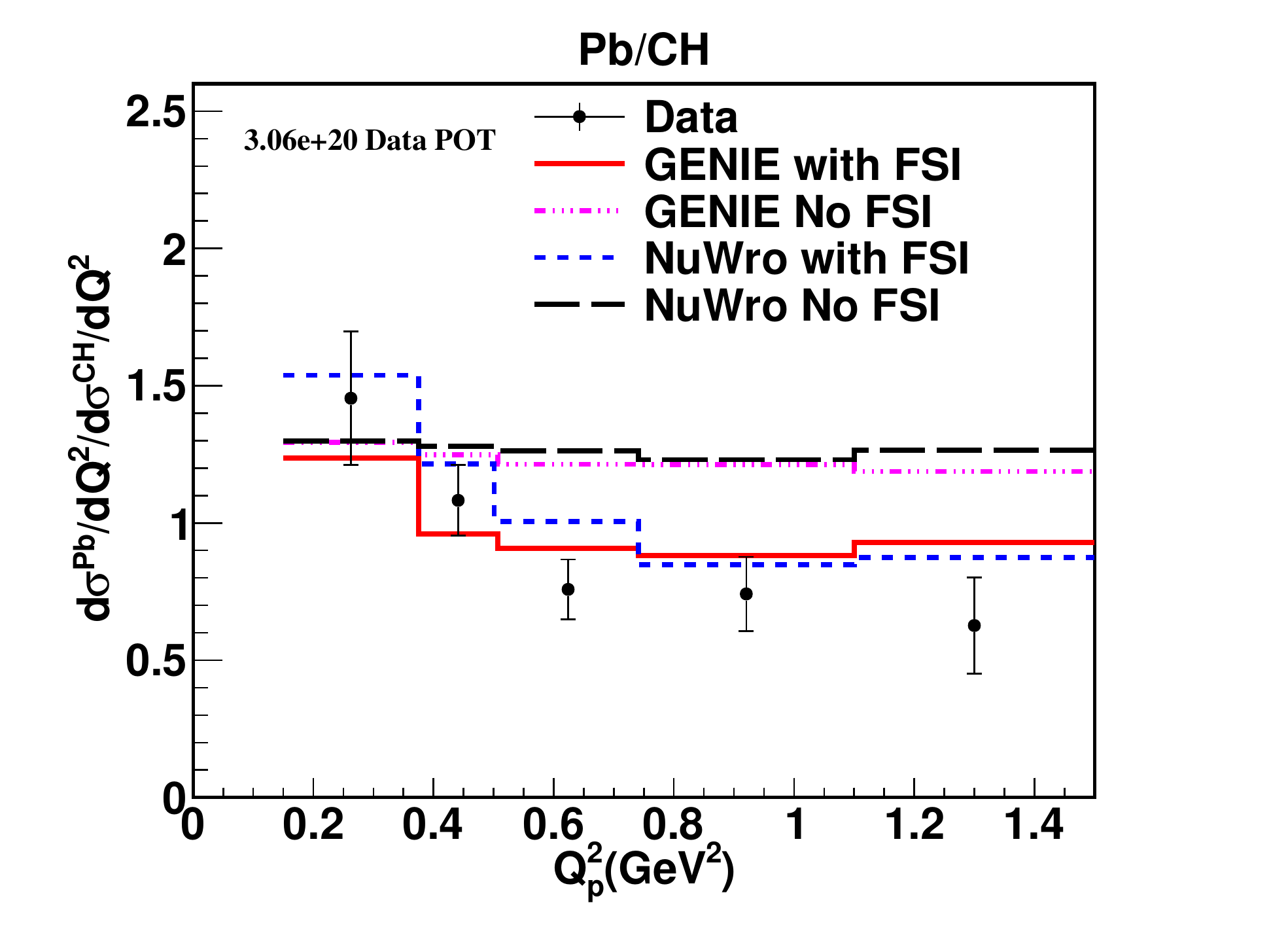}
\put (20,20) {(b) Pb/CH}
\end{overpic}
\caption{The \numu CCQE-like per-nucleon cross sections for (a) iron and (b) lead relative to CH  as a function of $Q^2$. \genie and \nuwro predictions with RPA and \tpth models (including the \tpth-like enhancement discussed in Sect.~\ref{sec:enhance}) are compared.  Figures from Ref.~\cite{Betancourt:2017uso}.}
\label{fig:nuclearqe}
\end{center}
\end{figure}

On the other hand, proton kinematics such as the proton-based $Q^2$ are sensitive to FSI whose strength depends on the size of the target nucleus due to the intranuclear energy loss. This motivated further quasielastic-like cross-section measurements by \minerva
using the detector's upstream target planes (carbon, iron, and lead) together with the active tracker (CH)~\cite{Betancourt:2017uso}. The extracted cross-section ratios of iron and lead to CH are shown in Fig.~\ref{fig:nuclearqe}. The data are found to be reasonably well described by
the $A$-dependent FSI prescriptions used by the \genie and \nuwro event generators.

\subsection{Initial-State Correlations and \tpth-Like Enhancement}\label{sec:enhance}

The nuclear response are often described theoretically using the energy transfer, $q_0$ (also called $\omega$ or $\nu$ in the literature), and the three-momentum transfer, $q_3$ or $\left|\vec{q}\right|$, from the lepton
to the target nucleus, whereby $q_3^2-q_0^2=Q^2$. However, the estimation of $q_0$ experimentally introduces uncertainties due to missing energy from nucleon unbinding and neutrons in the final state.  To avoid
systematic errors that may enter in this way,
\minerva 
introduced a new observable called ``available energy", \eav, as a proxy for $q_0$ in the ``low-recoil'' (meaning $q_3<\unit[0.8]{GeV}$) analyses using the Low-Energy \numu and \antinumu CC inclusive samples~\cite{Rodrigues:2015hik,Gran:2018fxa}.  Available energy is defined as the  
sum of the energies from final-state particles that can be calorimetrically measured in the scintillator and excludes removal energy and neutron energy. An estimate of $q_0$ is still used with muon kinematics to calculate \enu and then $q_3$, but its model uncertainties are subdominant for $q_3$.

\begin{figure}[!htb]
\begin{center}
\begin{overpic}[trim={0 1cm 0 0.59cm},clip,width=\doublewidth]{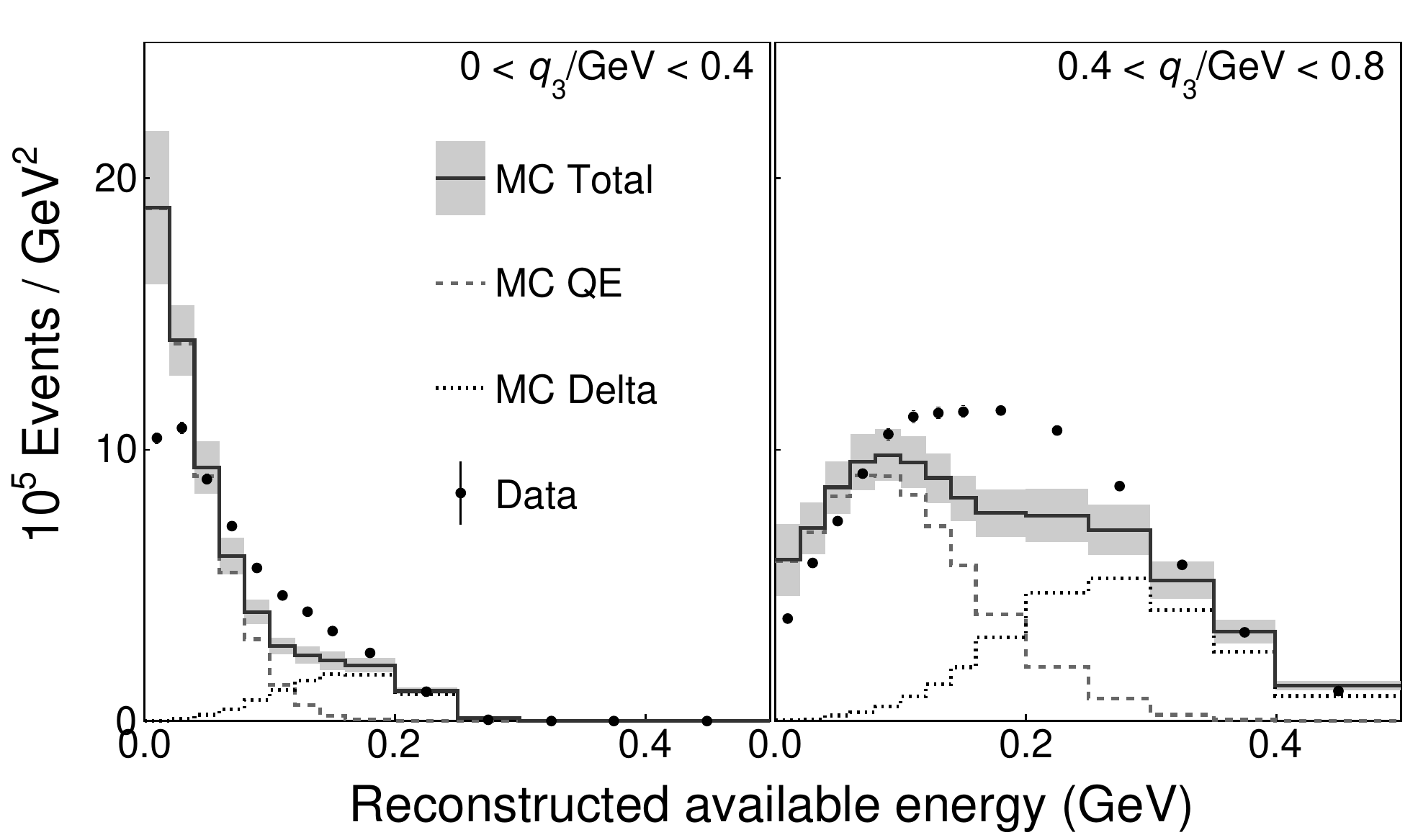}
\put (62,42) {\large (I) \numu, default}
\end{overpic}
\begin{overpic}[trim={0 1.6cm 0 0.59cm},clip,width=\doublewidth]{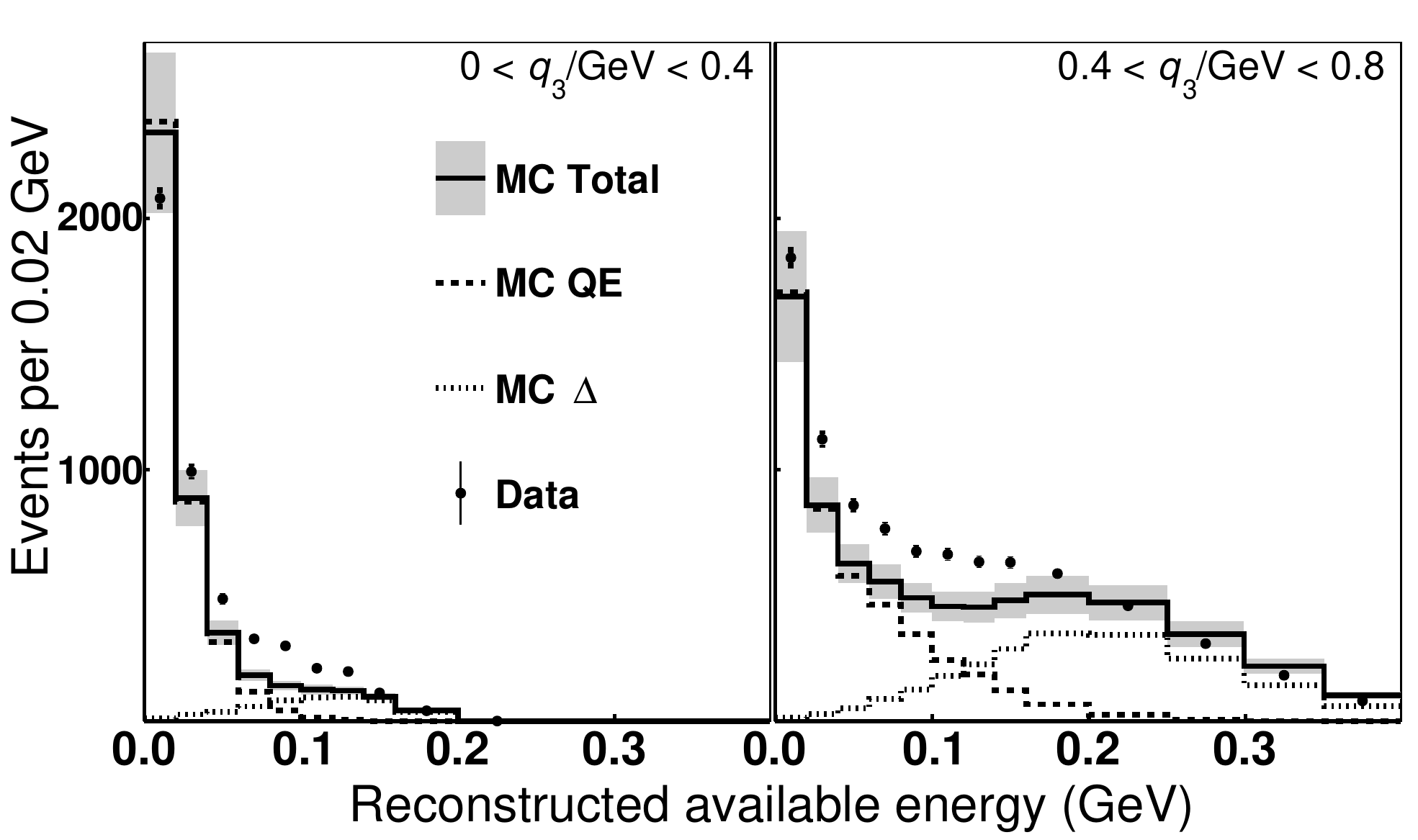}
\put (62,38) {\large (II.a) \antinumu, default}
\end{overpic}
\begin{overpic}[trim={0 0 0 0.59cm},clip,width=\doublewidth]{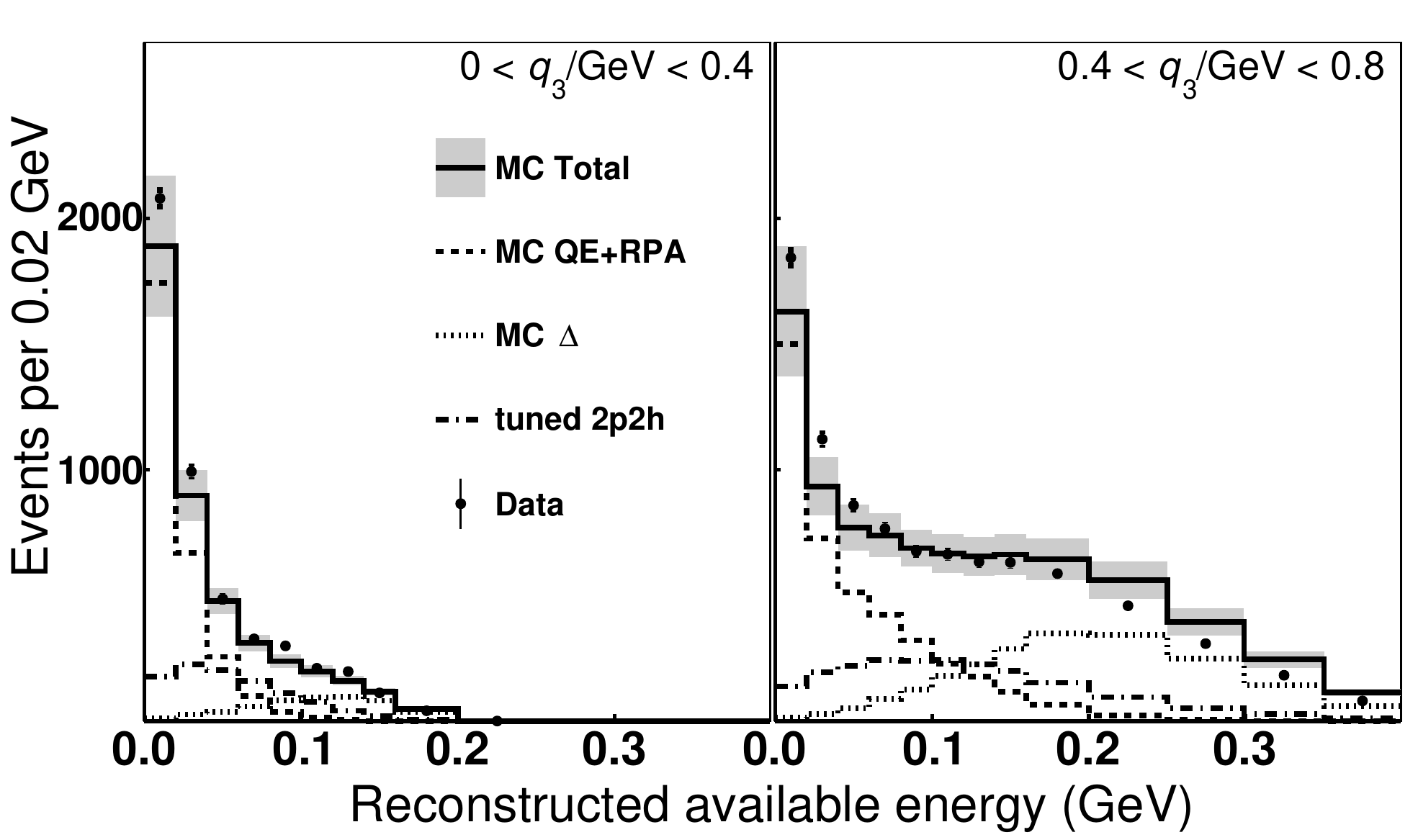}
\put (62,45) {\large (II.b) \antinumu, w/ tunes}
\end{overpic}
\caption{Event distributions in the reconstructed \eav for two ranges of the  reconstructed $q_3$ in (I) \numu and (II) \antinumu charged-current inclusive samples. In (I) and (II.a) the default simulations (see text) are used, while in (II.b) the predictions with the \numu-based tunes (\mnv) are shown. In the legend, the resonance region is labeled as Delta ($\Delta$). Figures from Refs.~\cite{Rodrigues:2015hik,Gran:2018fxa}.}
\label{fig:eav}
\end{center}
\end{figure}

The measured distributions for reconstructed \eav of \numu scattering~\cite{Rodrigues:2015hik} are shown in Fig.~\ref{fig:eav} (I) for two regions of the reconstructed $q_3$. 
They are compared to the default simulation based on \genie with modifications to pion production~\cite{Wilkinson:2014yfa,Rodrigues:2016xjj}. The simulation is further improved by taking into account initial-state correlations: the collective long-range medium effect in quasielastic events calculated with a Random Phase Approximation (RPA) approach~\cite{Nieves:2004wx}, and the Valencia QE-like \tpth model~\cite{Nieves:2011yp,Sobczyk:2012ms,Gran:2013kda,Schwehr:2016pvn}. Inclusion of RPA brings the event rates at low \eav into better agreement with the data, though beyond-Fermi-gas models may have a similar effect~\cite{Martini:2016eec,Nieves:2017lij}. The Valencia \tpth model helps reduce the model deficit at the dip region between quasielastic and resonance production. 
The best data-model agreement is achieved by separately scaling up the \tpth event rates in regions of $q_0$ and $q_3$: across all $q_0$-$q_3$ regions, an enhancement by \unit[50]{\%} is required to enable the model to
match the data, and in the dip region an enhancement of a factor of 2 is needed. 
Since this \textit{ad hoc} enhancement is based on the Valencia \tpth model, it was first interpreted as a correction to the modeled \tpth  mechanism. The \genie model that has evolved to this end is denoted as \mnv in subsequent \minerva publications.

The corresponding \antinumu measurement~\cite{Gran:2018fxa} is shown in Fig.~\ref{fig:eav} (II), where the aforementioned default simulations and \mnv are compared in (II.a) and (II.b), respectively. Importantly, the \tpth enhancement in \mnv is only tuned to the \numu measurement. Consequently, its predictive power for \antinumu, coming from the same underlying $q_0$-$q_3$-dependence as for \numu, is intriguing.  It suggests that the enhancement is at the level of structure functions  rather than of FSI or Fermi motion. However, given its empirical nature, this enhancement should be understood as  \tpth-\emph{like}; the analysis does not currently rule out a quasielastic or resonant enhancement in addition to \tpth.

\begin{figure}[!ht]
\begin{center}
\begin{overpic}[trim={0.6cm 0.7cm 0.8cm 1.1cm},clip,width=\singlewidth]{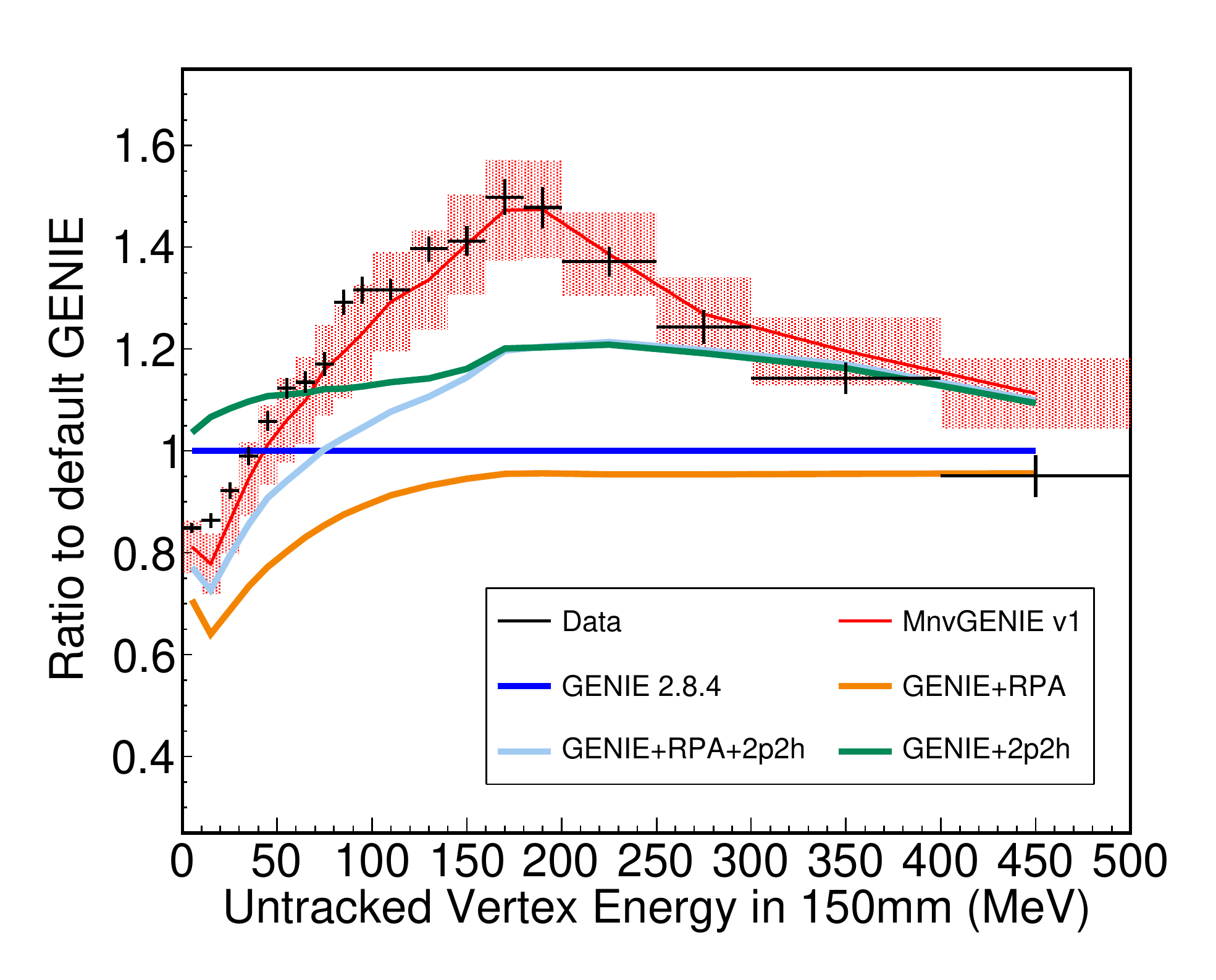}
\put (60,70) {(a) no $\proton$ tracks}
\end{overpic}
\begin{overpic}[trim={0.6cm 0.7cm 0.8cm 1.1cm},clip,width=\singlewidth]{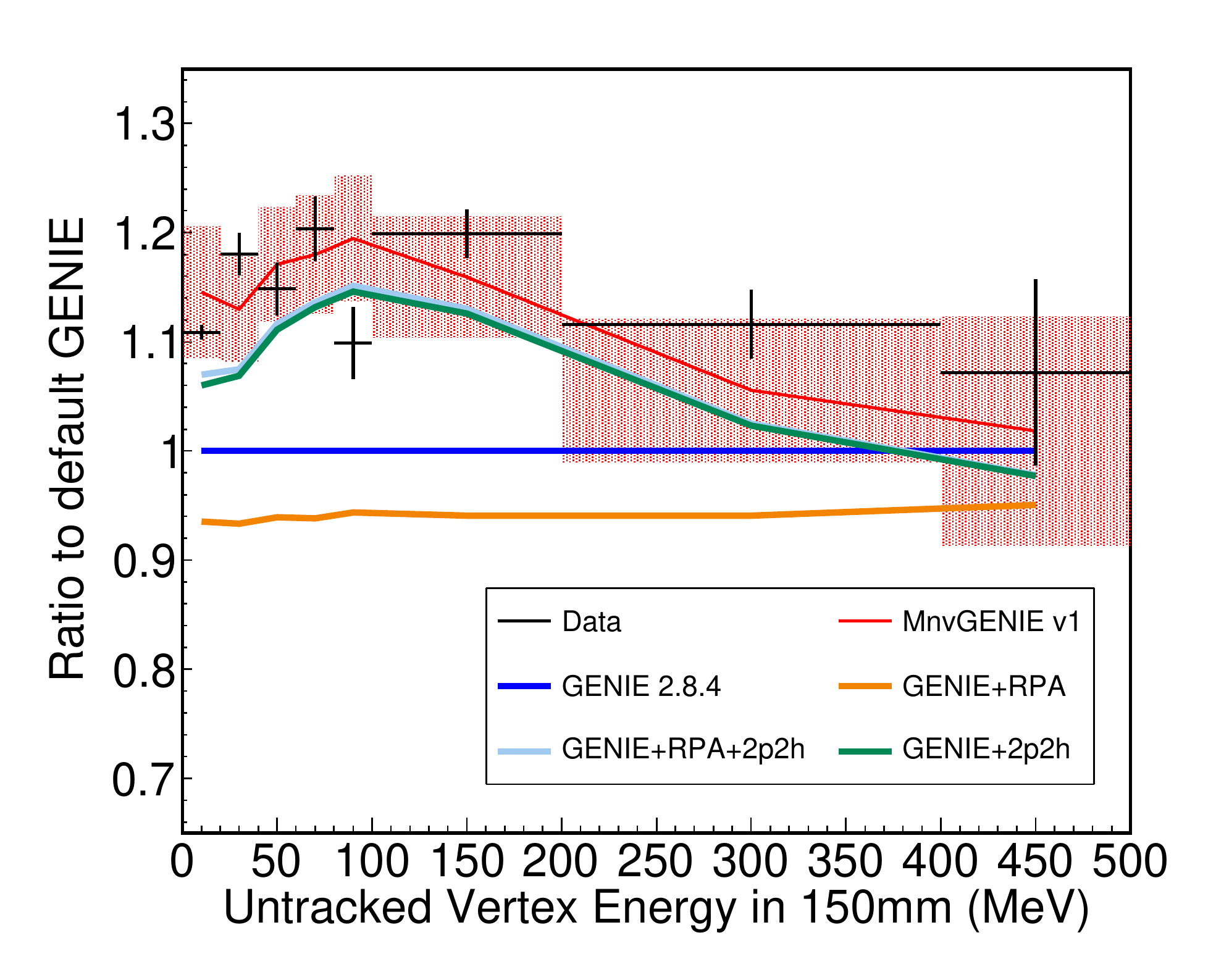}
\put (60,70) {(b) w/ $\proton$ tracks}
\end{overpic}
\caption{Ratio of the data and various \genie predictions to \genie 2.8.4 as a function of the reconstructed vertex energy in \numu CCQE-like events (a) without and (b) with proton tracks. The \tpth-like enhancement can be seen by comparing \mnv and ``\genie+RPA+(Valencia)\tpth''. Figures from Ref.~\cite{Ruterbories:2018gub}.}
\label{fig:vertexe}
\end{center}
\end{figure}

The inclusion of RPA and \tpth in 
 the predictions has made a significant improvement in the description of the \minerva data.
 The evolution of modeling improvement from \genie to \mnv is illustrated by a reanalysis of the vertex energy in a \numu CCQE-like sample~\cite{Ruterbories:2018gub}. As is shown in Fig.~\ref{fig:vertexe}, the distributions of the reconstructed vertex energies in events with and without proton tracks are described by \genie after the modifications with RPA, Valencia \tpth, and the \tpth-like enhancement are taken into account. 

\subsection{State-of-the-Art Quasielastic-Like Measurements}\label{sec:2ndqe}

\begin{figure}[t]
\begin{center}
\includegraphics[width=\triplewidth]{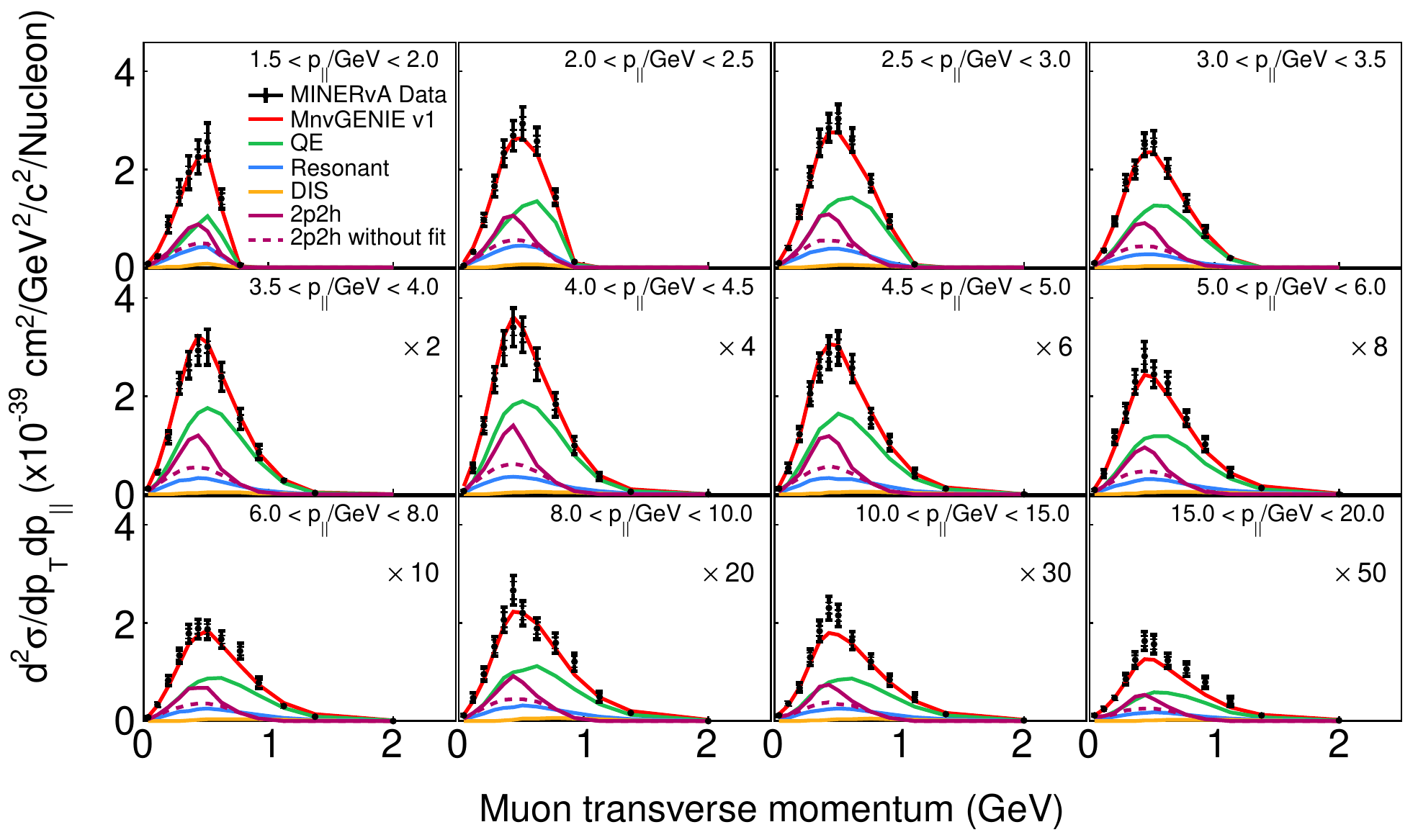}
\caption{\numu CCQE-like cross sections in \pt in bins of \pz. Unstacked curves show the various components of the \mnv predictions. ``\tpth without fit'' refers to the Valencia model while ``\tpth'' includes the \textit{ad hoc} enhancement discussed in Sect.~\ref{sec:enhance}. Figure from Ref.~\cite{Ruterbories:2018gub}.}
\label{fig:ptpz}
\end{center}
\end{figure}

In the few-GeV range of the neutrino energy, the quasielastic cross section 
is nearly constant as a function of \enu, while 
the phase space for resonance production and DIS is opening up. 
In addition to extracting the CCQE-like cross sections in $Q^2$ and \enu which can be  calculated with the quasielastic hypothesis (that is invalid for the non-QE components), \minerva measured the CC muon transverse (\pt ) and longitudinal momenta (\pz)  using the Low-Energy \antinumu~\cite{Patrick:2018gvi}, \numu~\cite{Ruterbories:2018gub}, and Medium-Energy \numu~\cite{Carneiro:2019jds} data sets. These particular muon momentum projections respectively
approximate the true $Q^2$ and \enu.  As shown in Fig.~\ref{fig:ptpz} for the Low-Energy \numu results, the different interaction contributions in the \mnv predictions are relatively stable across the \pz bins.  This is in contrast to the inclusive measurement discussed in Sect.~\ref{sec:inelastic} below where the \pz- (and therefore, \enu-) dependence of the DIS processes is evident. In both the QE-like and the inclusive measurements, while \mnv describes the data in most bins, there is an overall model deficit at large \pz.

\begin{figure}[!ht]
\begin{center}
\begin{overpic}[trim={0.6cm 0 0.6cm 0},clip,width=\halfwidth]{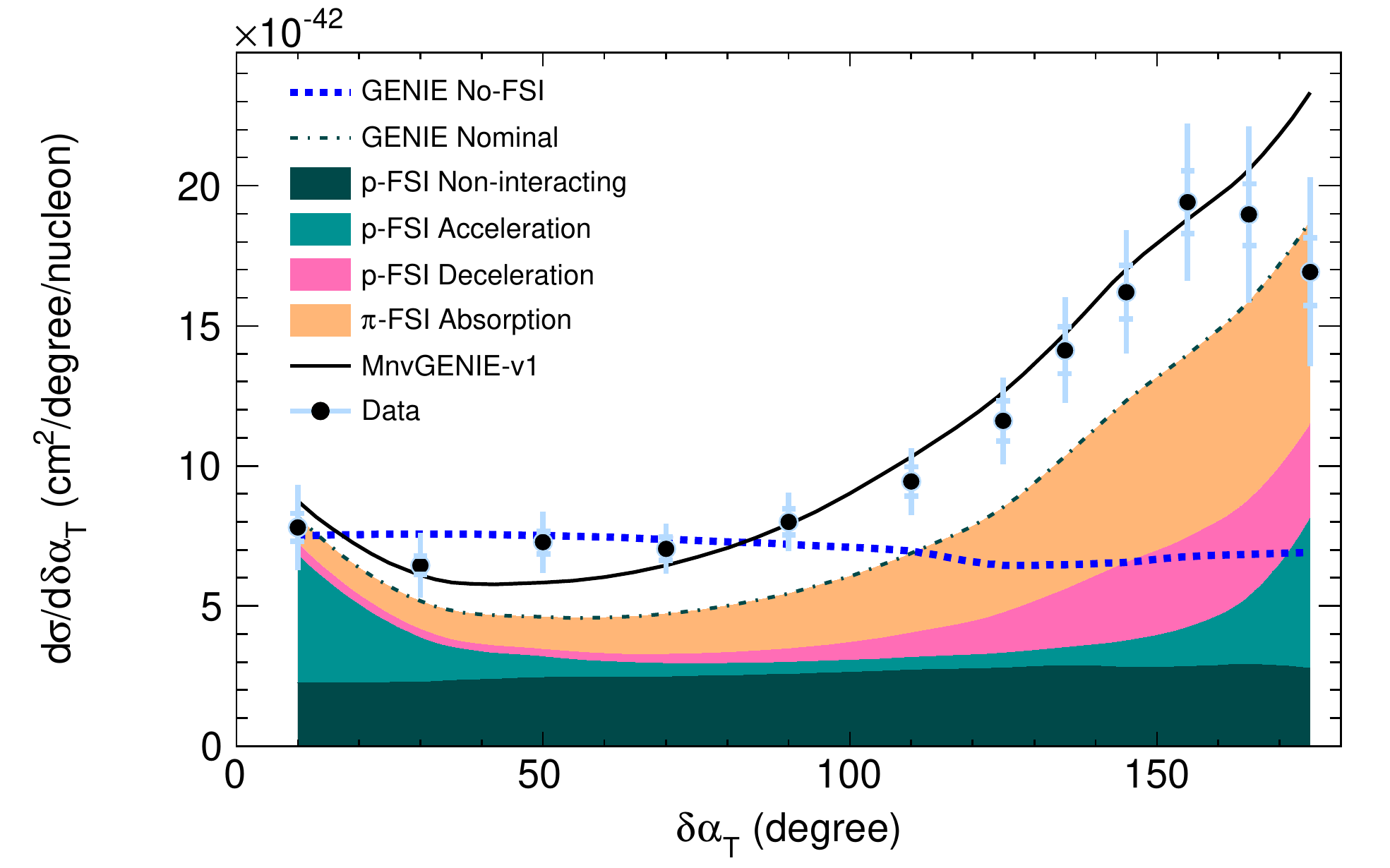}
\put (55,50) {(a)}
\end{overpic}
\begin{overpic}[trim={0.6cm 0 0.6cm 0},clip,width=\halfwidth]{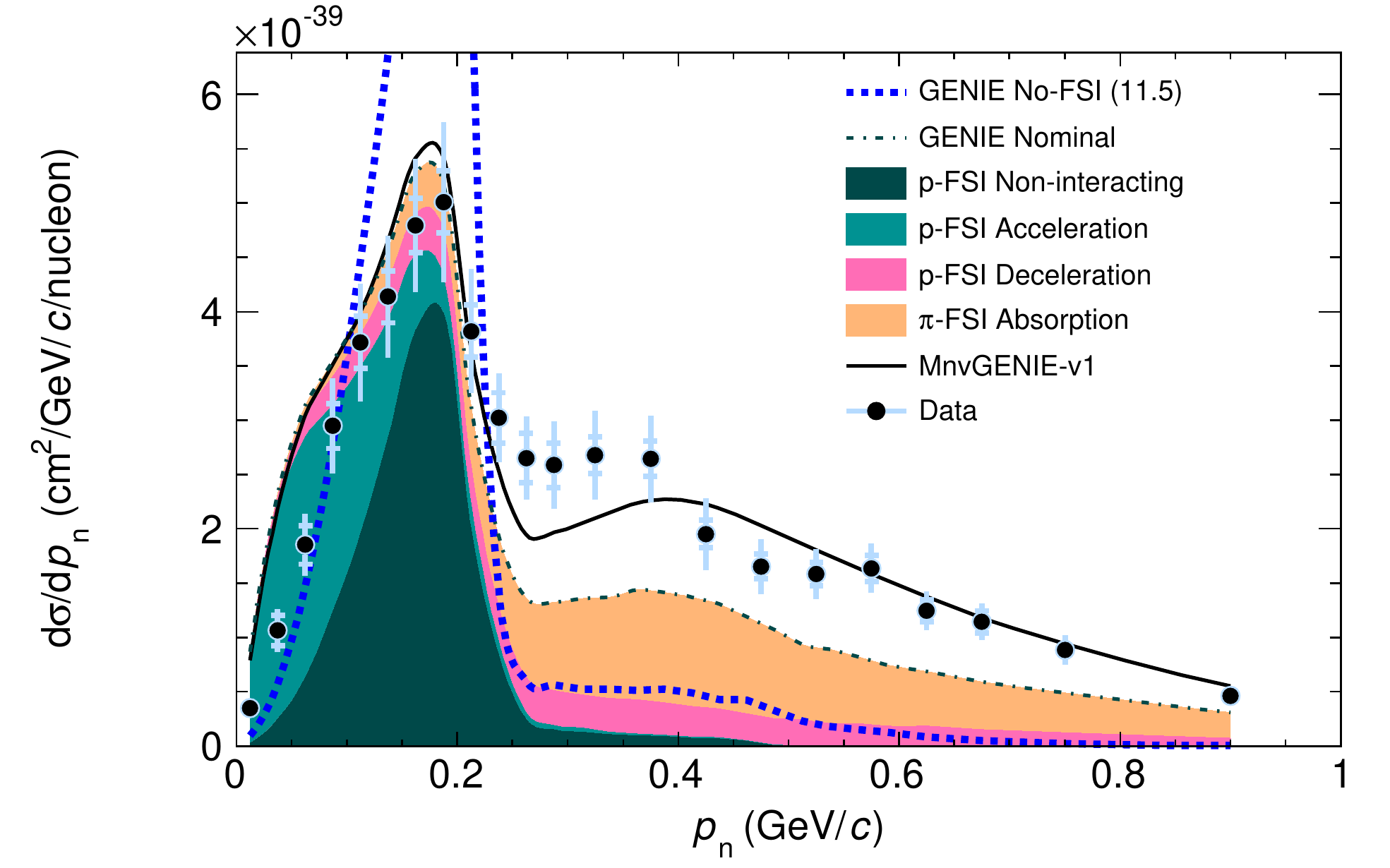}
\put (45,50) {(b)}
\end{overpic}
\caption{Differential cross sections for \numu CCQE-like events in (a) \dat and (b) \pn. The stacked colored histograms depict categories of FSI and sum up to the \genie nominal predictions. The predicted \tpth component including the \tpth-like enhancement can be inferred from the difference between \mnv and the nominal \genie. Figures from Ref.~\cite{Lu:2018stk}.}
\label{fig:qetki}
\end{center}
\end{figure}

By measuring the transverse kinematic imbalance which cancel out the primary interaction kinematics, the cross section dependence on the incoming neutrino energy is lessened and also the initial-and final-state effects can be directly probed~\cite{Lu:2015tcr}. 
Using the \numu CCQE-like events in the tracker with the Low-Energy data~\cite{Lu:2018stk}, the direction and magnitude of the transverse momentum imbalance (\vdpt) between the muon and the leading proton, \dat and \dpt, respectively, are calculated. The angle \dat has the most sensitivity to FSI and to the unaccounted-for momentum carried by missing particles such as absorbed pions or the correlated nucleon of the proton from \tpth. Figure~\ref{fig:qetki} shows that, within the uncertainties, \mnv describes the data. Here, FSI are classified into three categories: 
\begin{enumerate}
    \item A flat distribution for events which do not experience FSI (both the ``no-FSI'' and ``p-FSI non-interacting'' categories in Fig.~\ref{fig:qetki}) reflecting the isotropy of the Fermi motion; 
    \item The deceleration region (\dat$\rightarrow180^\circ$) for energy-dissipating processes---decelerating FSI, pion absorption, and \tpth;
    \item The acceleration region (\dat$\rightarrow0^\circ$) for accelerating FSI if such a mechanism exists.
\end{enumerate}
Interestingly, \genie did predict FSI acceleration for protons, roughly half of them singularly occupying the acceleration region, while the other half falls into the deceleration region due to the transverse projection.

By assuming a carbon-11 target remnant, the transverse momentum imbalance magnitude \dpt is promoted to the three-dimensional momentum imbalance, \pn, following from an additional constraint by energy conservation~\cite{Furmanski:2016wqo}. For the FSI-noninteracting events, \pn can be interpreted as the momentum of the struck neutron in the CCQE initial state. The location of the Fermi-motion peak in data is well captured by the \mnv prediction. However, its accelerating FSI component causes the predicted peak shape to deviate from data. This component was identified as the elastic component of the \genie v2.8 \textit{hA} FSI model and has been removed in later versions of \genie~\cite{Harewood:2019rzy,Cai:2019hpx}. The measured cross section is further compared to \nuwro predictions with alternative nuclear states: local Fermi gas (LFG) and Spectral Function. The latter model better describes the Fermi motion peak, but neither model provides enough strength in the transition region between the quasielastic peak and the non-QE tail~\cite{Lu:2018stk}. 

\begin{figure}[!ht]
\begin{center}
\includegraphics[trim={0 0 0 0.5cm},clip,width=\singlewidth]{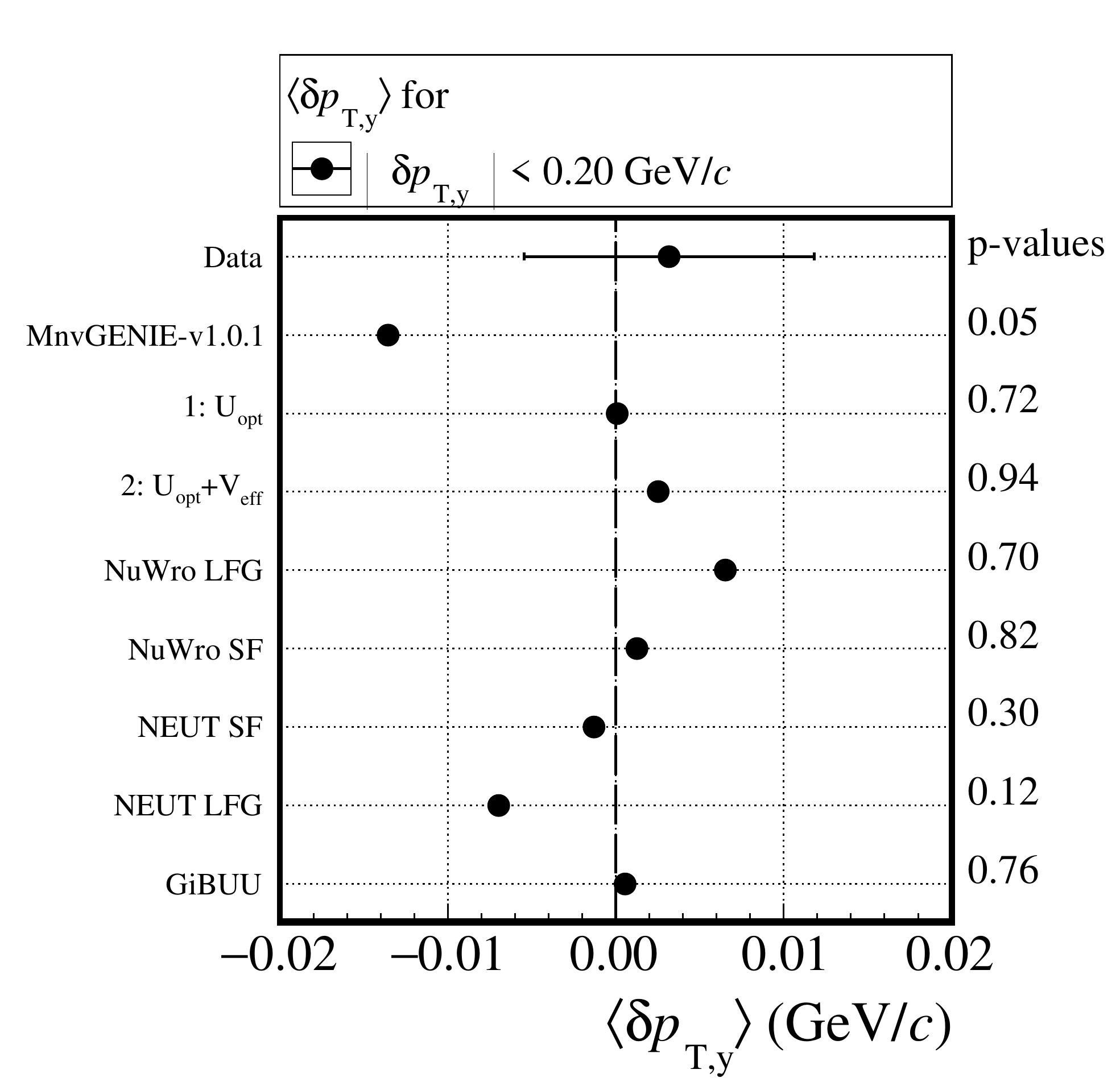}
\caption{Mean \dpty calculated from the \numu CCQE-like differential cross section in \dpty 
within \unit[-0.20 to 0.20]{\gevc}. The $p$-value is the probability,
assuming normal distribution, that the observed result would have
been produced by the test model. $U_\textrm{opt}$ and $V_\textrm{eff}$ are the optical and Coulomb potentials experienced by the CCQE proton and muon, respectively, as proposed in Ref.~\cite{Bodek:2018lmc}. \neut~\cite{Hayato:2009zz} and \gibuu~\cite{Buss:2011mx} predictions are also compared.  Figure from Ref.~\cite{Cai:2019hpx}.}
\label{fig:dpty}
\end{center}
\end{figure}

An enhanced sensitivity to another initial-state condition, the binding energy, is achieved by further projecting \vdpt onto the lepton scattering plane, which defines the \dpty variable~\cite{Cai:2019hpx}. Various generator implementations of the interaction energy on carbon are compared (Fig.~\ref{fig:dpty}) and the data favor approximate corrections to \genie~\cite{Bodek:2018lmc}.  

\subsection{Charged-Current Pion Production}\label{sec:pion}

Neutrino-induced pion production is an important channel in neutrino oscillation experiments because it accounts for a large part of the signal in \nova and \dune~\cite{Abi:2020wmh}  far detector event samples, and is both a low-statistics signal and background in the far detector samples of T2K.  The process can proceed through baryon resonance production or through non-resonant interaction.  It can also occur through a CC coherent interaction with a nucleus---see discussions in Sect.~\ref{cohpika}. Because of the additional final-state hadrons, incoherent pion production may be more affected by the nuclear environment than the quasielastic process, and the difficulty with reconstructing pions means that the process is harder to measure. One would like to use the lessons learned about nuclear effects in quasielastic-like scattering to (overt) pion production, but the effects may not be the same.  The nuclear responses to the pion production fall into three categories according to their relation with those of the quasielastic-like processes:
\begin{enumerate}
    \item Nuclear responses that are generic but might be quantitatively different in pion production are Fermi motion, binding energy, initial-state correlations, and Pauli blocking~\cite{Paschos:2003qr,Bodek:2020wbk}; 
    \item   Pion absorption and charge exchange are nuclear responses that migrate primary pion production channels among each other and into quasielastic-like topology~\cite{Salcedo:1987md,Merenyi:1992gf};
    \item Some nuclear responses are specific to pion production such as $\Delta$-resonance in-medium modifications~\cite{Oset:1987re,Lalakulich:2012cj,Hernandez:2013jka}.
\end{enumerate}
Moreover, in the energy region of \minerva, the contribution of higher resonances could be important along with the nuclear medium effects in the pion production processes. \minerva has measured the following processes on its plastic scintillator tracker with the Low-Energy beam:  
\begin{itemize}
    \item \numu CC $\pi^+$ production (with limited contributions from $\pi^-$)~\cite{Eberly:2014mra,McGivern:2016bwh},
     \item \numu CC single $\pi^0$ production~\cite{Altinok:2017xua,Coplowe:2020yea},
    \item \antinumu CC single $\pi^0$ production~\cite{Aliaga:2015wva,McGivern:2016bwh}, and
    \item \antinumu CC single $\pi^-$ production~\cite{Le:2019jfy}.
\end{itemize}   

\begin{figure}[b]
\begin{center}
\includegraphics[trim={0.4cm 0.2cm 0.5cm 0.9cm},clip,width=\halfwidth]{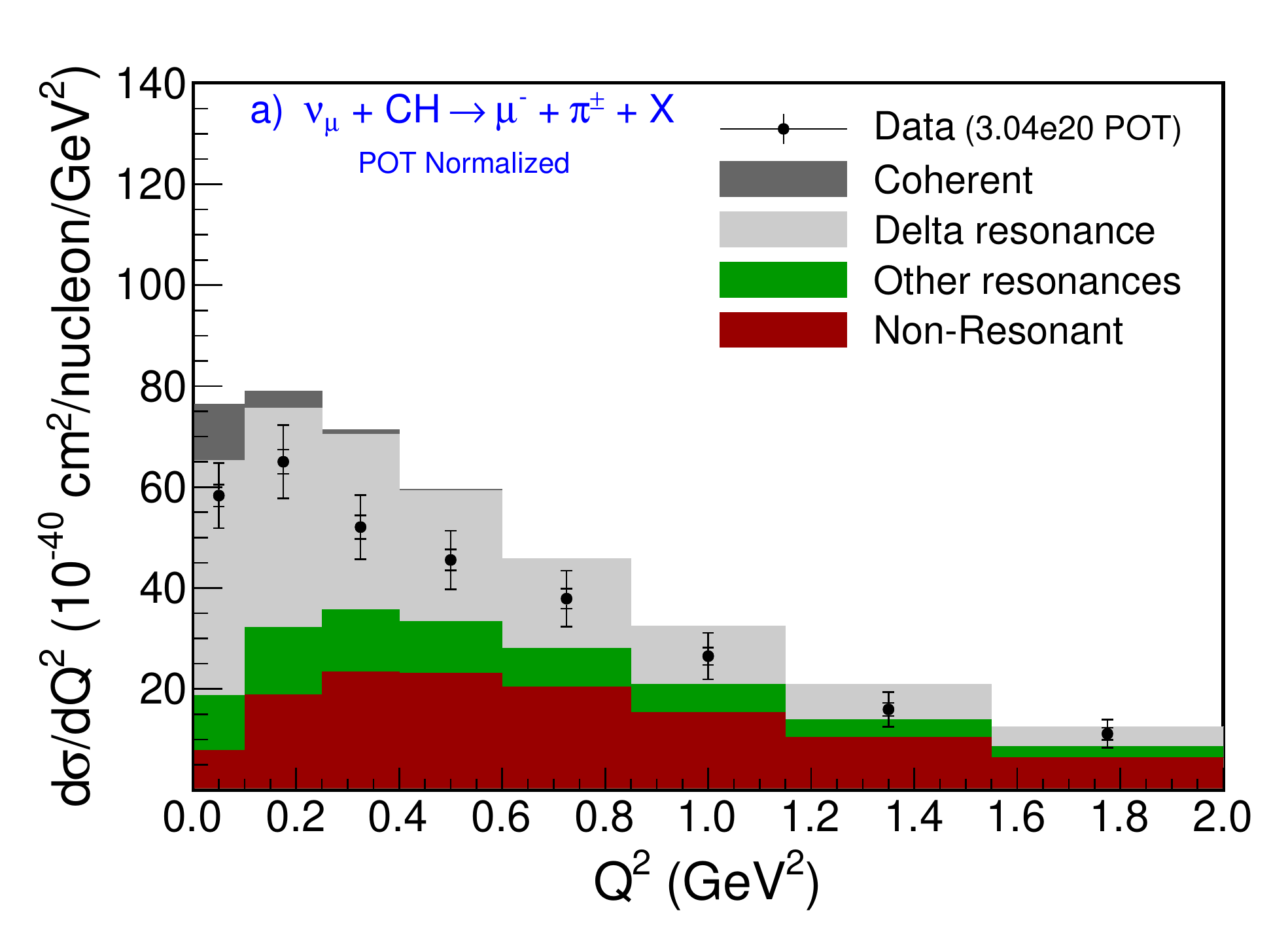}
\includegraphics[trim={0.4cm 0.2cm 0.5cm 0.9cm},clip,width=\halfwidth]{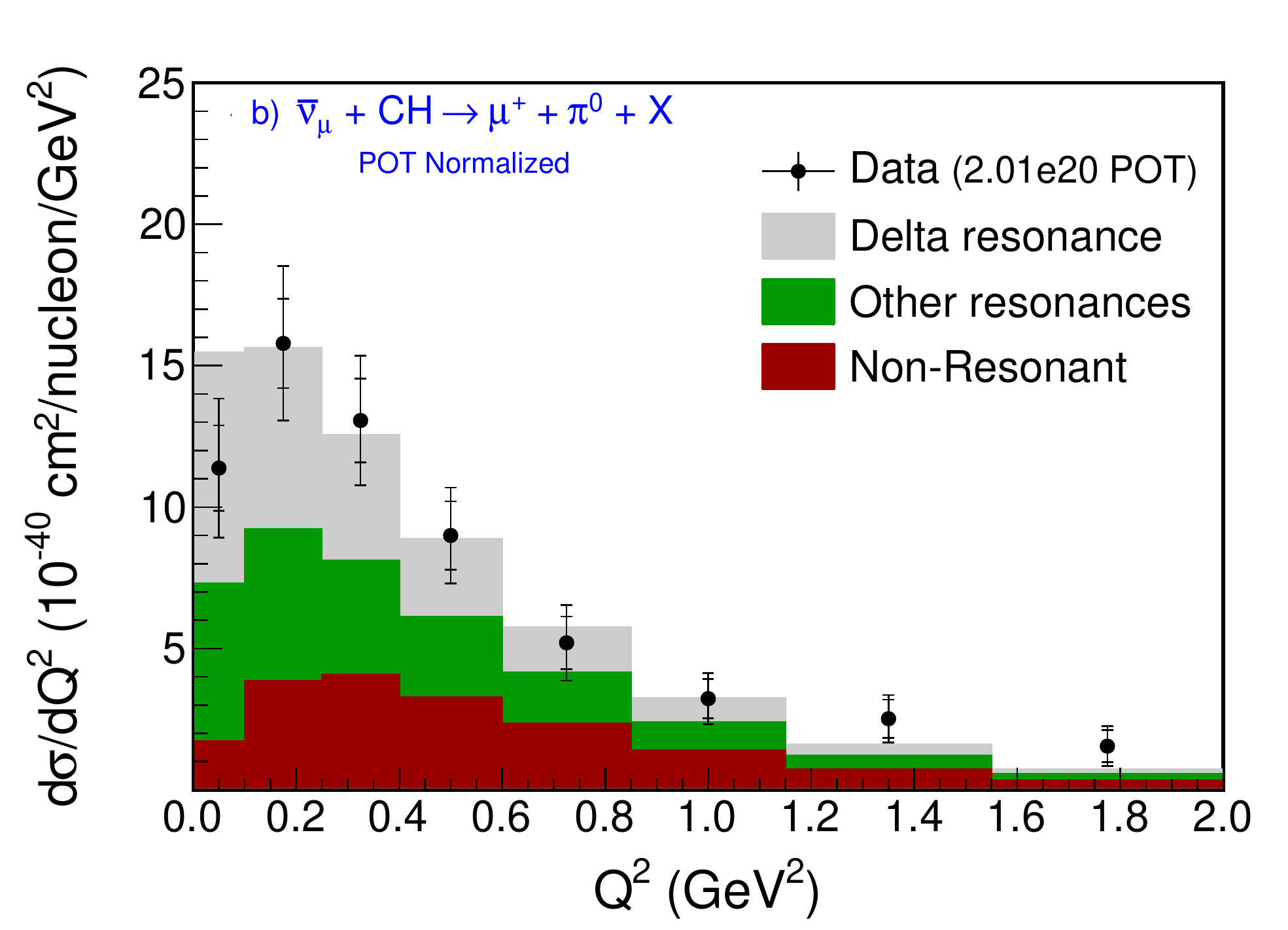}
\begin{overpic}[trim={0.4cm 0.2cm 0.5cm 0.9cm},clip,width=\halfwidth]{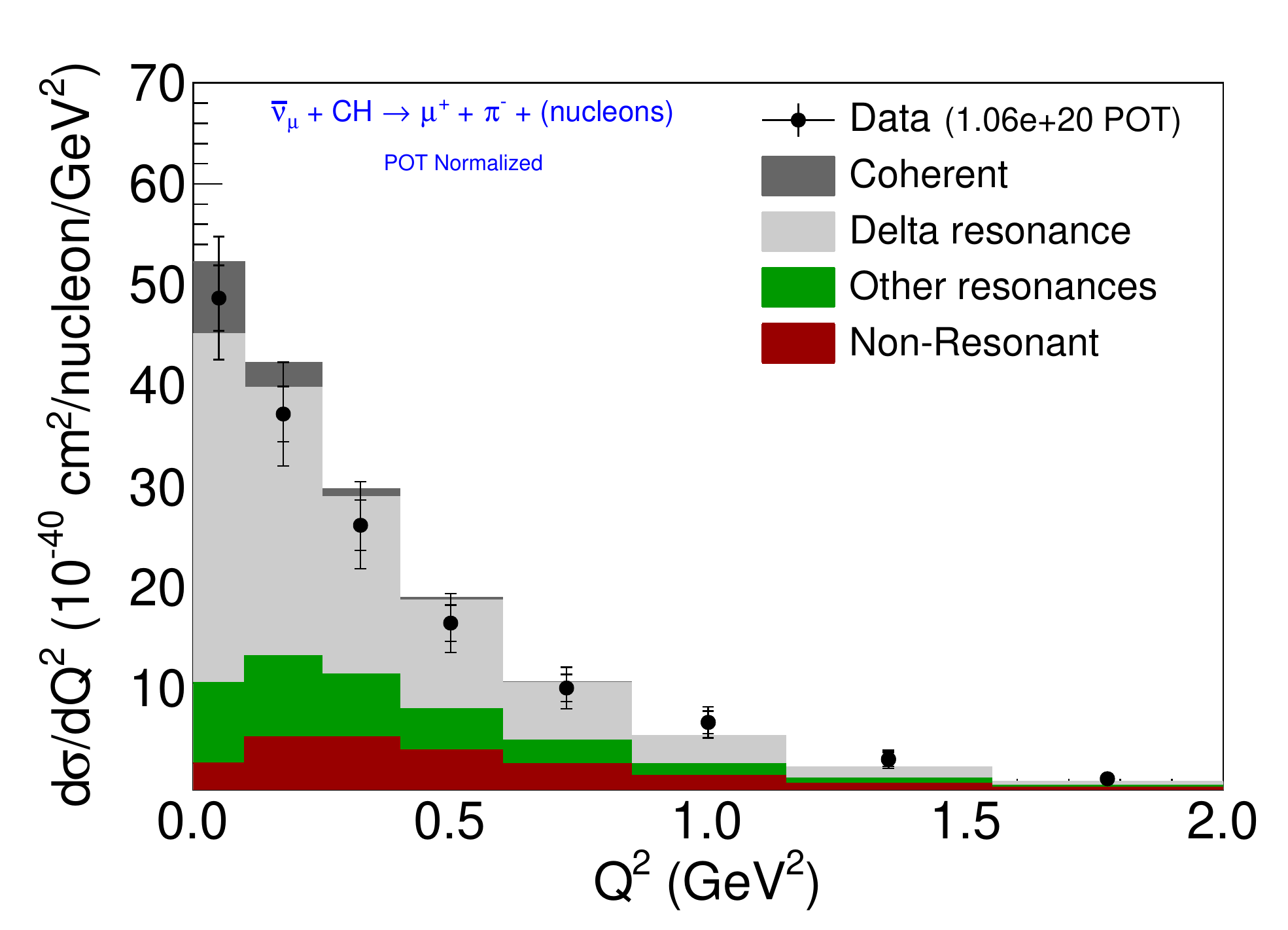}
\put (14.5,64.5) {\tiny \textcolor{blue}{(c)}}
\end{overpic}
\begin{overpic}[trim={0.4cm 0.2cm 0.5cm 0.9cm},clip,width=\halfwidth]{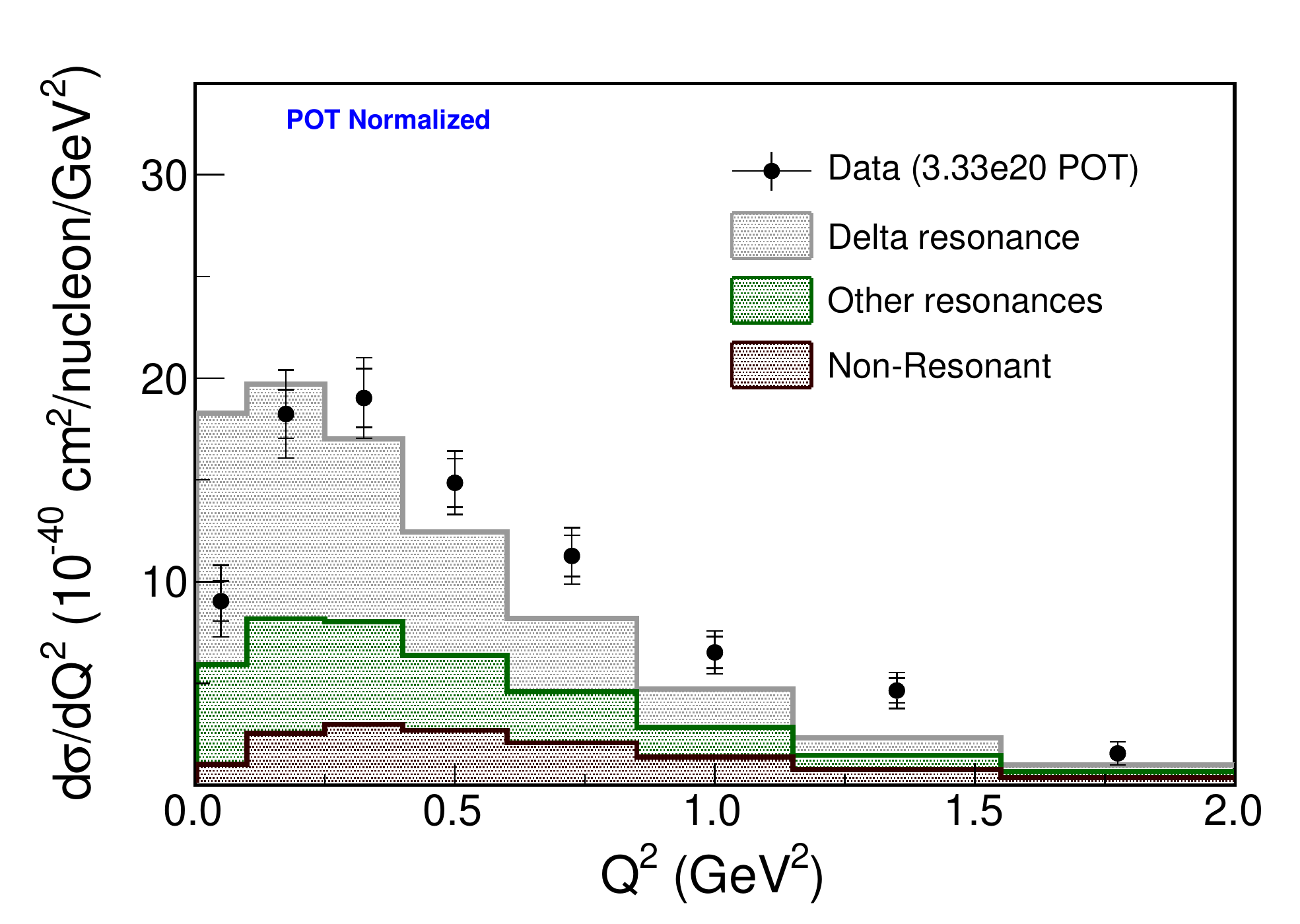}
\put (39,63) {\tiny \textcolor{blue}{(d) $\nu_\mu$+CH$\rightarrow\mu^-$+$\pi^0$+nucleon(s)}}
\end{overpic}
\caption{Cross sections in $Q^2$ for (a) \numu $\pi^+$ (with limited contributions from $\pi^-$), (b) \antinumu $\pi^0$, (c) \antinumu $\pi^-$, and (d) \numu $\pi^0$ charged-current productions compared to \genie predictions. Figures from Refs.~\cite{McGivern:2016bwh,Le:2019jfy,Altinok:2017xua}.}
\label{fig:piq2}
\end{center}
\end{figure}

The nuclear effects described by RPA and \tpth are crucial to the description of quasielastic-like processes in \minerva (Sect.~\ref{sec:enhance}).  Since the RPA effect creates a suppression of the QE-like cross section at low $Q^2$, measuring $Q^2$ in pion production could provide relevant information. However, as shown by the results in Fig.~\ref{fig:piq2}, 
suppression at low $Q^2$ ranges from being nonexistent (\antinumu $\pi^-$~\cite{Le:2019jfy}), to mild or insignificant (\numu $\pi^+$~\cite{Eberly:2014mra,McGivern:2016bwh} and \antinumu $\pi^0$~\cite{Aliaga:2015wva,McGivern:2016bwh}), and to fairly pronounced (\numu $\pi^0$~\cite{Altinok:2017xua}). 
A combined fit of the various underlying mechanisms using a subset of these measurements can be found in Ref.~\cite{Stowell:2019zsh}.  
On the other hand, while \tpth in pion production has not been incorporated into predictions, current models describe the \minerva data with sufficient strength without it. The effect of initial-state correlations on pion production is not well understood.

\begin{figure}[!ht]
\begin{center}
\includegraphics[trim={0.25cm 0 0.25cm 0},clip,width=\halfwidth]{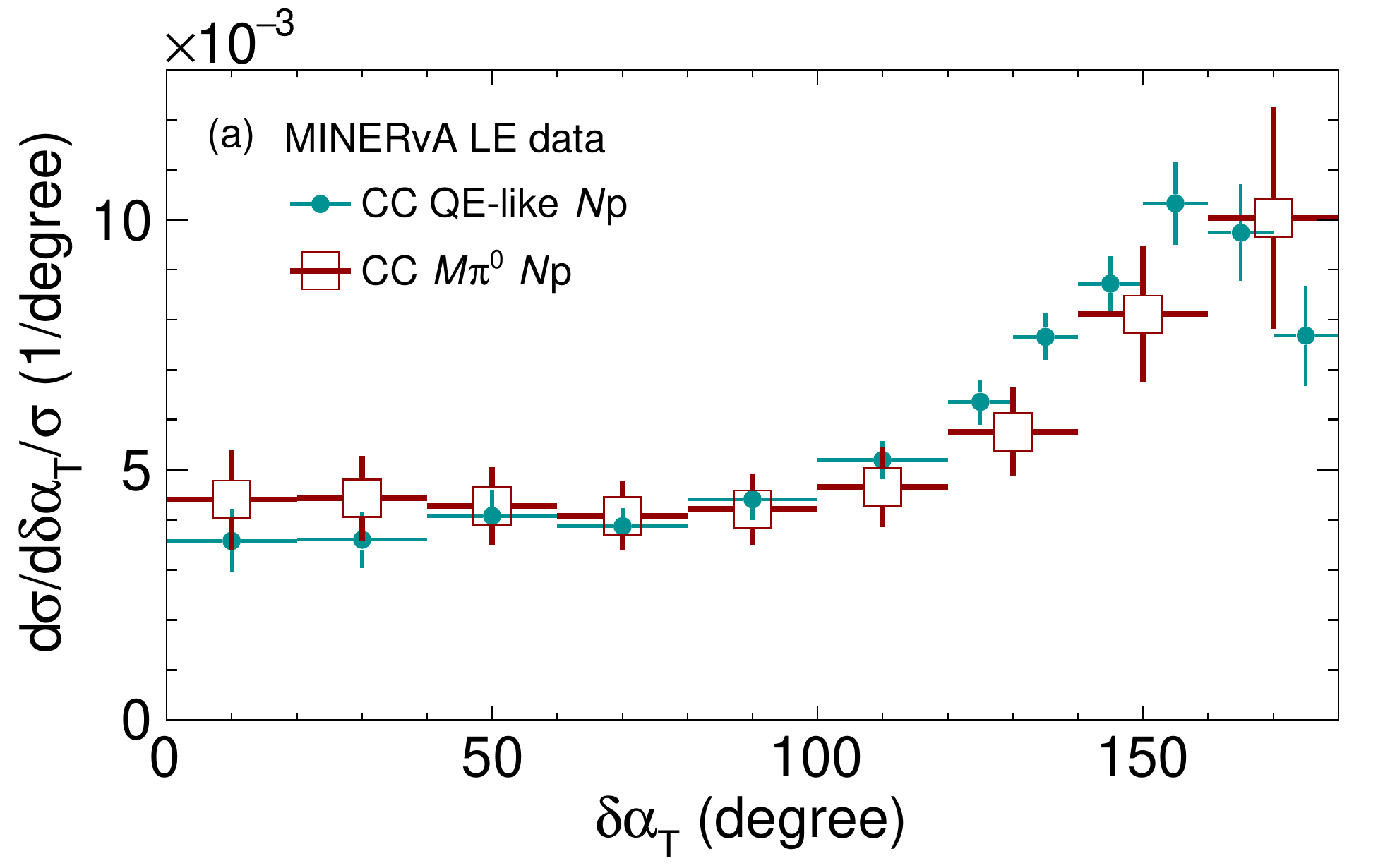}
\includegraphics[trim={0.25cm 0 0.25cm 0},clip,width=\halfwidth]{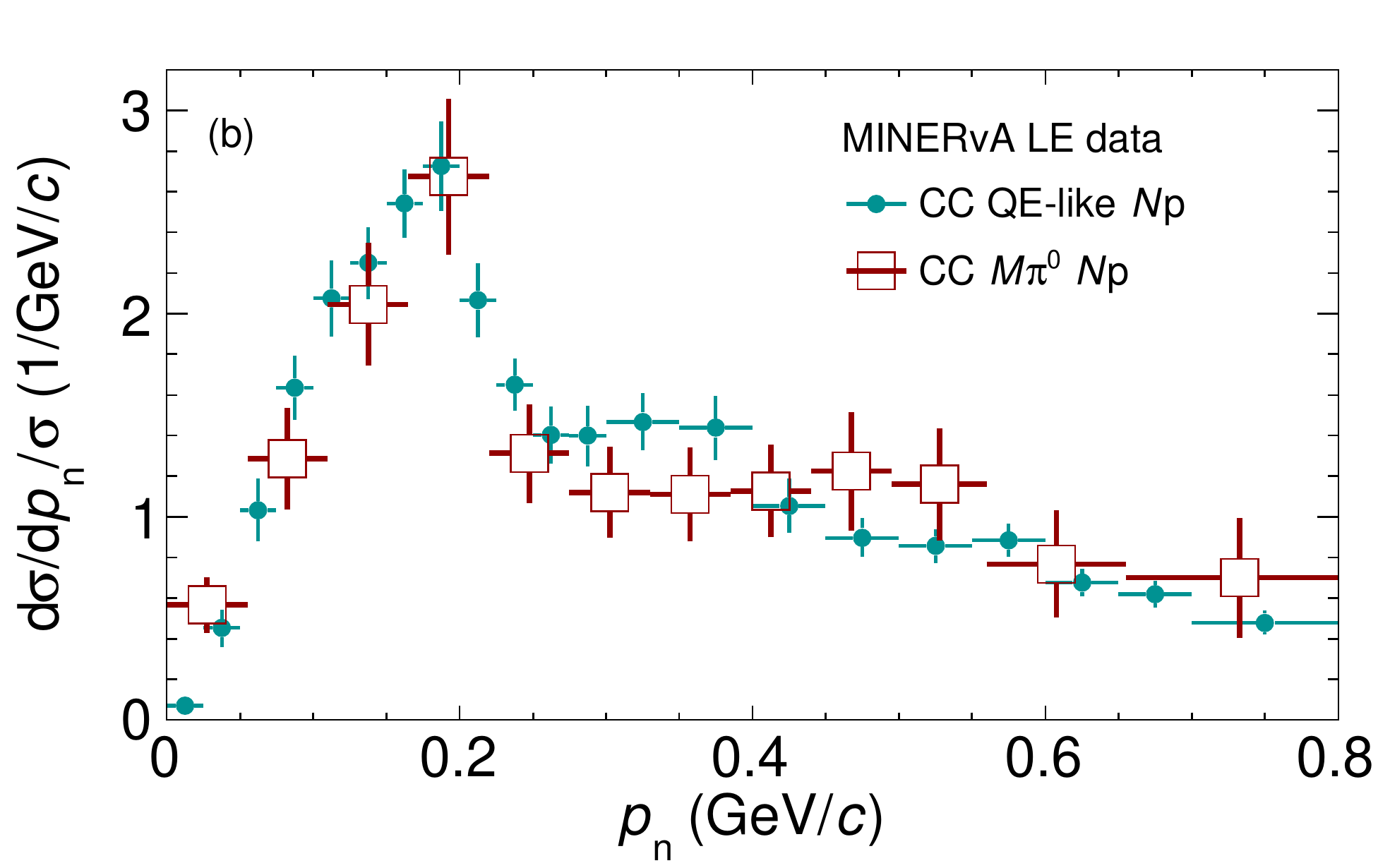}
\caption{Differential cross sections in (a) \dat and (b) \pn for
\numu charged-current $\proton$-$\pi^0$ production, compared to quasielastic-like scattering~\cite{Lu:2018stk,Cai:2019hpx} (Sect.~\ref{sec:2ndqe}). The data sets are area-normalized. Figures from Ref.~\cite{Coplowe:2020yea}.}
\label{fig:piontki}
\end{center}
\end{figure}

Because of the granularity of the tracker and the event statistics obtained for \numu charged-current proton-$\pi^0$ final states with the Low-Energy beam, \minerva is able to investigate the transverse kinematic imbalances that may arise in pion production~\cite{Coplowe:2020yea}.  
In this way, \minerva can probe the initial state and FSI in parallel to the CCQE-like measurement discussed in Sect.~\ref{sec:2ndqe} (Fig.~\ref{fig:piontki}). The TKI between the muon and the $\proton$-$\pi^0$ hadronic system is calculated by reconstructing the $\pi^0$ momentum and combining it with the proton momentum~\cite{Lu:2019nmf}. 
As with \numu CCQE, this channel also has an initial-state neutron and as expected, the Fermi motion peaks in Fig.~\ref{fig:piontki} (b) from both channels are consistent. The consistency in the \pn tail size and in the trend of \dat in Fig.~\ref{fig:piontki} is purely coincidental.  As a specific example, if nature had less pion absorption, the QE-like \pn tail and the \dat deceleration region would fall but the overall $\pi^0$ production would increase.  Currently, for the same Fermi motion peak, generator model predictions do not describe both channels simultaneously~\cite{Coplowe:2020yea}, clearly illustrating the challenges inherent to consistent modeling of few-GeV \nuA interactions. 

\subsection{Kaon Production}\label{sec:kaon}

\begin{figure}[!ht]
\begin{center}
\begin{overpic}[trim={0 0 2.4cm 0.6cm},clip,width=\halfwidth]{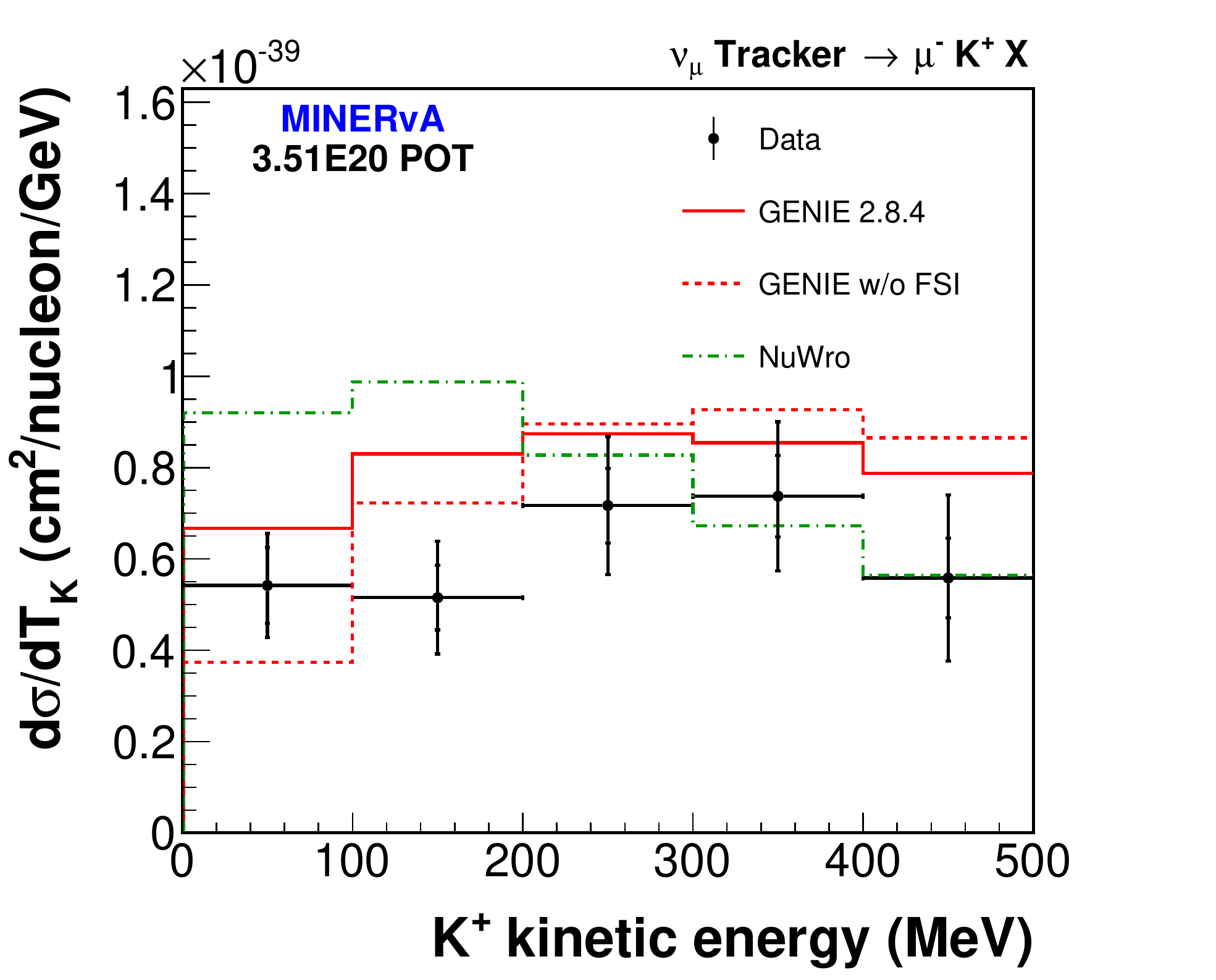}
\put (23,68) {(a) CC}
\end{overpic}
\begin{overpic}[trim={0 0 2.4cm 0.6cm},clip,width=\halfwidth]{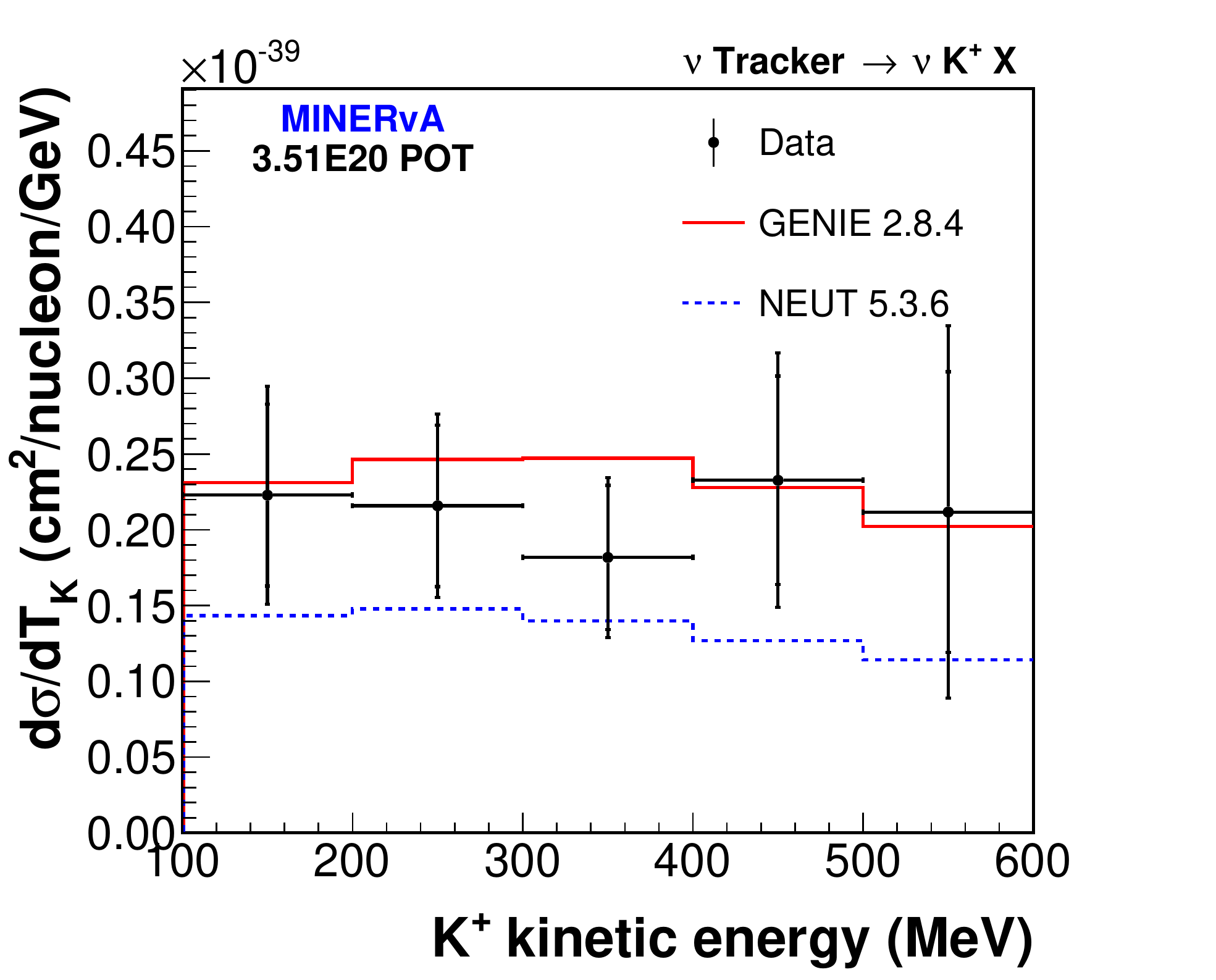}
\put (23,68) {(b) NC}
\end{overpic}
\caption{Charged-current (a) and neutral-current (b) \numu \kplus production cross sections in kinetic energy,
compared to \genie, \nuwro, and \neut predictions. Figures from Refs.~\cite{Marshall:2016rrn,Marshall:2016yho}.}
\label{fig:kaon}
\end{center}
\end{figure}

Charged and neutral kaons are produced in both charged-current and neutral-current interactions at the beam energies of the \minerva exposures. Kaon production is important to measure because the neutral-current (NC) channels yielding \kplus mesons constitute a background to proton decay searches~\cite{Abe:2014mwa,An:2015jdp}.  This background arises from interactions of atmospheric neutrinos that are incident on proton decay detectors.  
The production mechanism is associated production with another strange meson or with a hyperon. 
Below the threshold of associated particle production, kaon production would take place through $\Delta S=1$ currents ($S$ is the strangeness quantum number). At the energies of present interest, single kaon production may be important~\cite{RafiAlam:2010kf}. 
A final-state \kplus with energy less then $\sim$\unit[600]{MeV} can stop inside the \minerva tracker and then decay at rest.  Once a \kplus  decay chain is identified through the decay-associated time delay, the event topology, and the energy loss, the  \kplus initial momentum can be calculated from track range. For neutral-current interactions, the final-state lepton does not identify the vertex location, consequently the 100-MeV tracking threshold sets a lower limit on the measured kaon kinetic energy.  For CC interactions on the other hand, the final-state muon determines the primary vertex as the \kplus starting point, hence the kaon kinetic energy threshold can be lower since a kaon track needs not be identified. Figure~\ref{fig:kaon} shows the measured charged-current~\cite{Marshall:2016rrn} and neutral-current~\cite{Marshall:2016yho} cross sections. The analysis showed that the \genie prediction was in agreement with the neutral-current measurement, while the \neut prediction, which was used by Super-K~\cite{Abe:2014mwa}, predicted a lower cross section by about \unit[20]{\%}.

\subsection{Inelastic Reactions}\label{sec:inelastic}

In quasielastic and resonance production, the dynamic degrees of freedom are baryons and mesons. The invariant mass of the hadronic system, $W$, is of the order of \unit[1-2]{\gevcc}. In the high-$W$ high-$Q^2$ region where DIS dominates, the target nucleon breaks up in the reaction and the QCD dynamics in the nuclear environment can be studied. This kinematic region is only accessible in inclusive measurements where any hadronic final state is allowed. 

\begin{figure}[!ht] 
\begin{center}
\includegraphics[width=\linewidth]{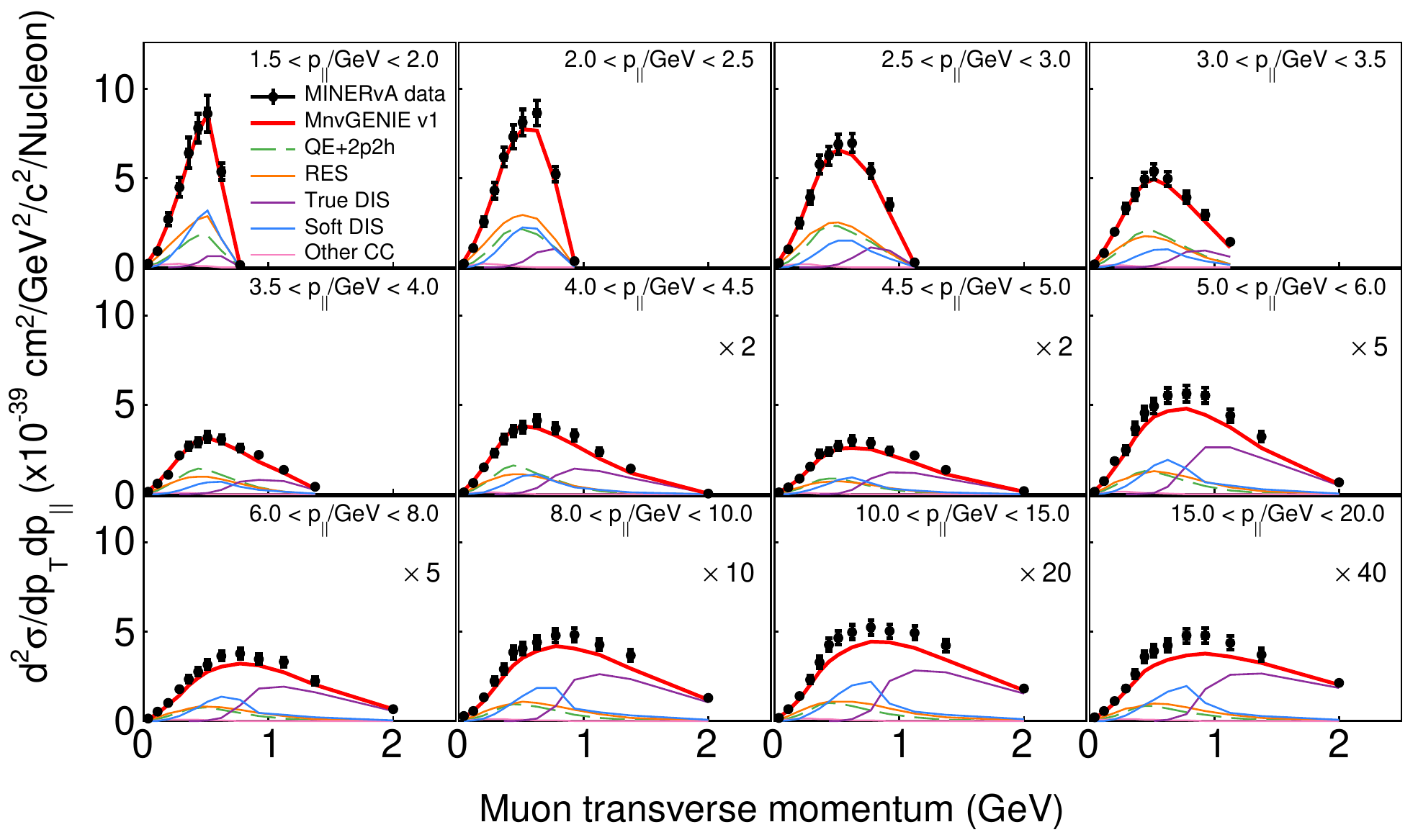}
\caption{\numu charged-current inclusive cross sections in \pt in bins of \pz compared to \mnv predictions with unstacked contributions shown.  See Fig.~\ref{fig:ptpz} for comparison. ``True DIS'' events are defined as \genie DIS events that satisfy $W >$ \unit[2]{\gevcc} and $Q^2 >$ \unit[1]{GeV$^2$}. ``Soft DIS'' is defined to contain the remaining \genie DIS events. Figure from Ref.~\cite{Filkins:2020xol}.}
\label{fig:inc2d}
\end{center}
\end{figure}

\minerva measures CC inclusive cross sections using the muon kinematics.  
These measurements require one momentum-analyzed muon per event; no requirement is placed on the final-state hadronic system~\cite{Filkins:2020xol}. Figure~\ref{fig:inc2d} shows
the \pz-evolution of the inclusive \pt spectra. In contrast to the CCQE-like measurement of  Fig.~\ref{fig:ptpz} of Sect.~\ref{sec:2ndqe}, there are significant contributions from (\genie) DIS in all \pz bins and its relative contribution increases with \pz. As in the CCQE-like case, there is an overall deficit of \mnv at large \pz.

\begin{figure}[t]
\begin{center}
\begin{overpic}[trim={0 0 2.9cm 0.8cm},clip,width=\halfwidth]{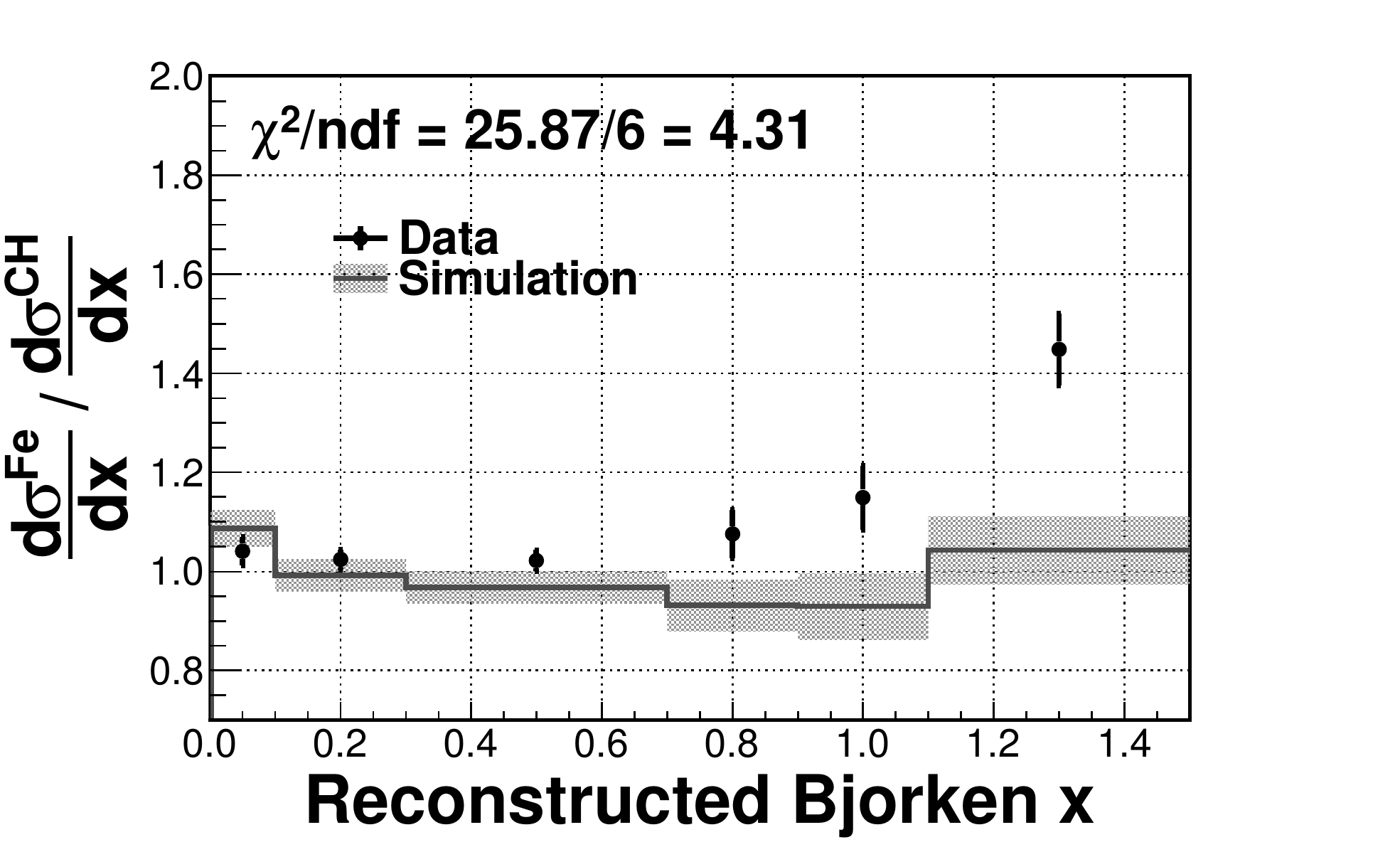}
\put (30,34) {(a) Fe/CH}
\end{overpic}
\begin{overpic}[trim={0 0 2.9cm 0.8cm},clip,width=\halfwidth]{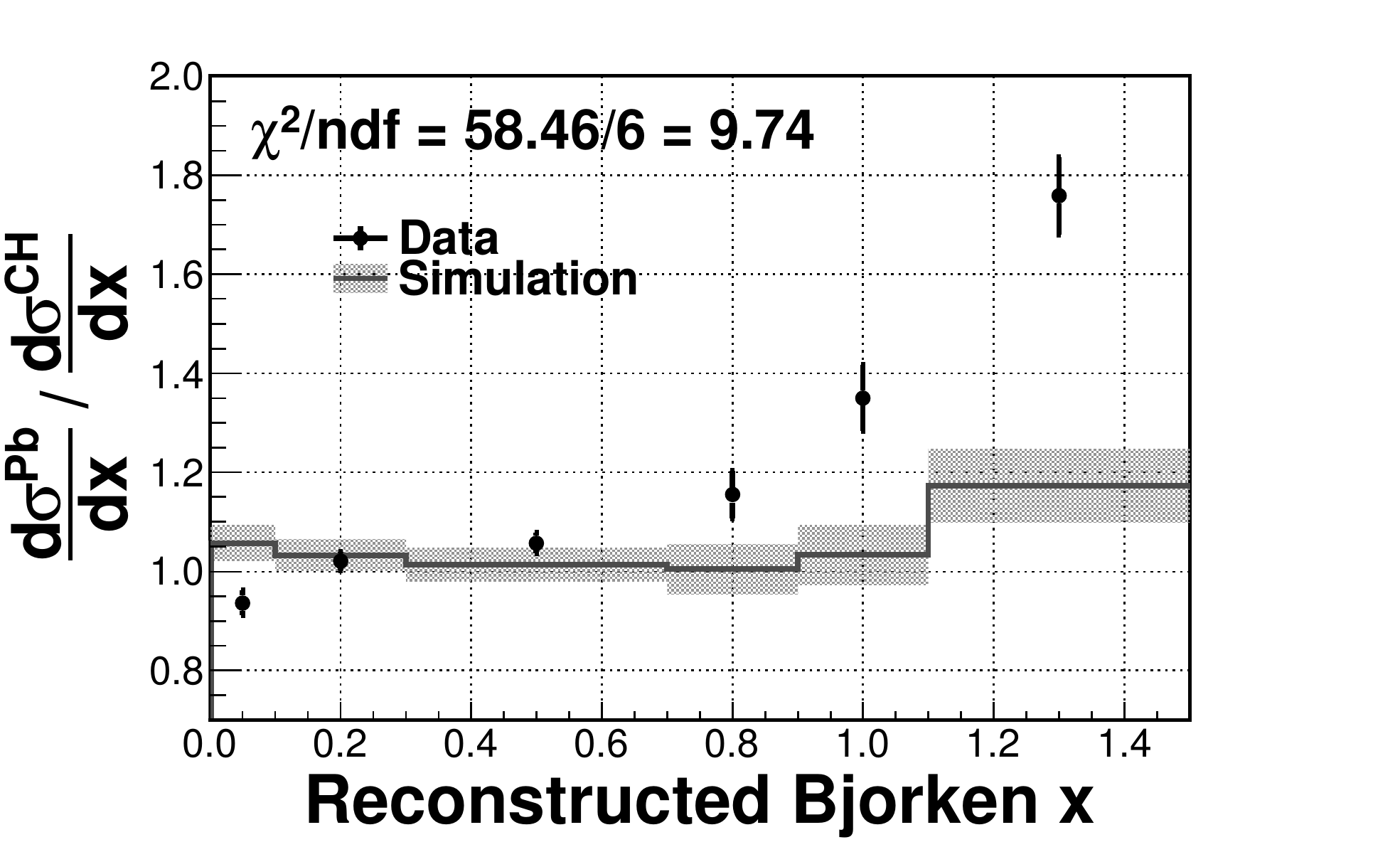}
\put (30,34) {(b) Pb/CH}
\end{overpic}
\caption{Ratios of \numu charged-current inclusive cross sections per nucleon as a function of the  reconstructed \bjx for (a) Fe/CH and (b) Pb/CH, compared to the default \genie simulations. Figures from Ref.~\cite{Tice:2014pgu}.}
\label{fig:Bjinc}
\end{center}
\end{figure}

Charged-current inclusive cross sections have also been measured in the passive target region as a function of the reconstructed \bjx.  These measurements are carried out for iron and lead targets; they are based on both muon kinematics and  calorimetric hadronic energy~\cite{Tice:2014pgu}. The ratios of \numu-Fe and \numu-Pb cross section to the cross section in the tracker (CH) are shown in Fig.~\ref{fig:Bjinc}. The default \genie simulations show a strong deficit in the elastic region (\bjx$\sim1$) which might be due to the unmodeled \tpth contributions.

\begin{figure}[t]
\begin{center}
\begin{overpic}[trim={0 0 2.9cm 0.8cm},clip,width=\halfwidth]{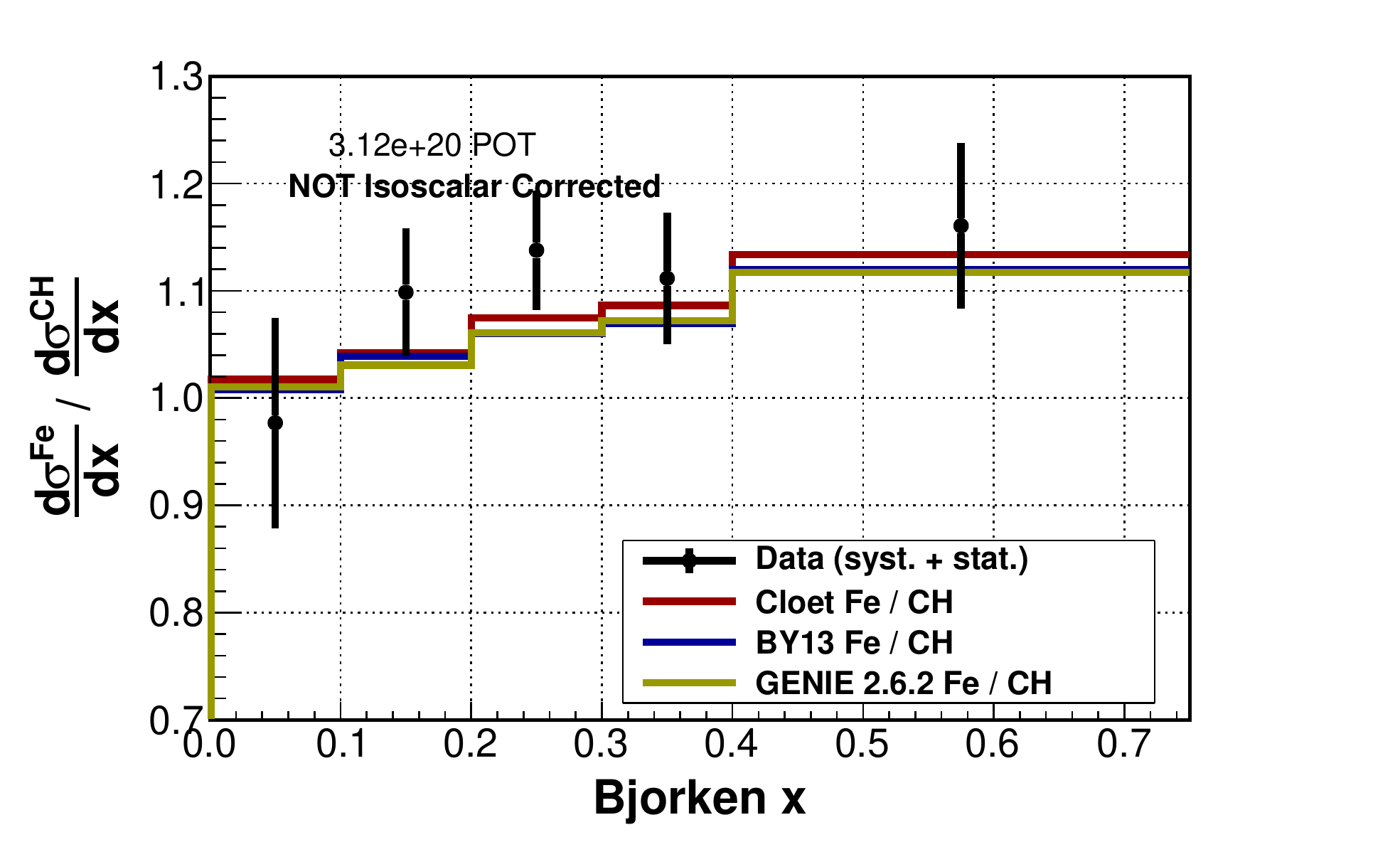}
\put (64,32) {(a) Fe/CH}
\end{overpic}
\begin{overpic}[trim={0 0 2.9cm 0.8cm},clip,width=\halfwidth]{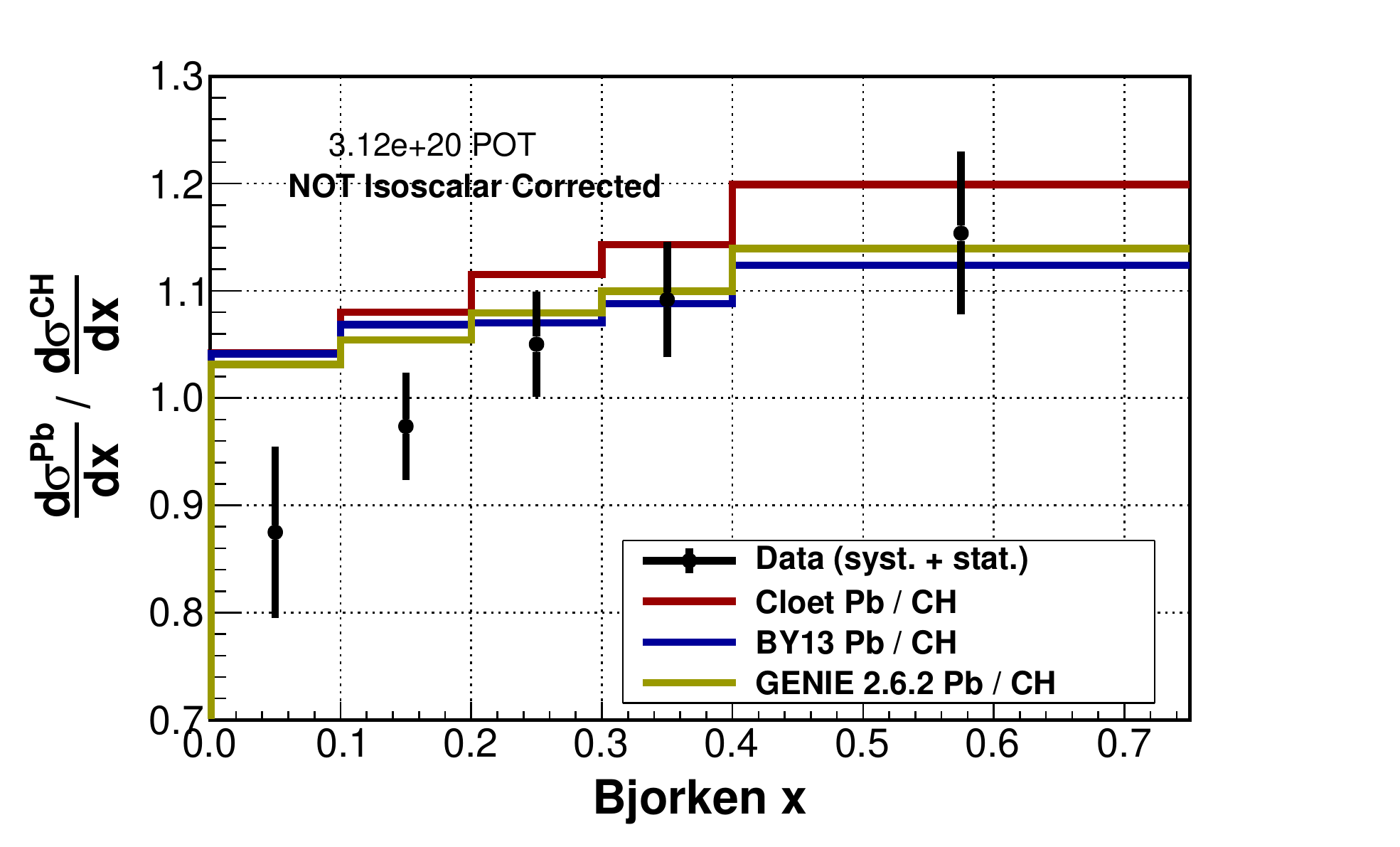}
\put (64,32) {(b) Pb/CH}
\end{overpic}
\caption{Ratios of \numu charged-current DIS cross sections per nucleon as a function of \bjx for (a) Fe/CH and (b) Pb/CH, compared to the default \genie, Bodek-Yang (BY13)~\cite{Bodek:2010km}, and Cloet~\cite{Cloet:2006bq} models. Figures from Ref.~\cite{Mousseau:2016snl}.}
\label{fig:BjDIS}
\end{center}
\end{figure}

By restricting the sample to the DIS region, $W >$ \unit[2]{\gevcc} and $Q^2 >$ \unit[1]{GeV$^2$}, these \bjx-dependent cross section ratios can be compared to additional DIS models~\cite{Mousseau:2016snl} (Fig.~\ref{fig:BjDIS}). As is shown in Fig.~\ref{fig:BjDIS} (b) for Pb, the data at \bjx$<0.1$ suggests possible shadowing effects beyond those predicted. Because these models are only tuned to charged-lepton scattering data, which are by definition insensitive to the axial-vector current, these data also reflect neutrino-specific effects in the DIS region. 

\section{Coherent Interactions}\label{cohpika}

\begin{figure}[b]
\begin{center}
\begin{overpic}[trim={0 0 2.2cm 0},clip,width=\halfwidth]{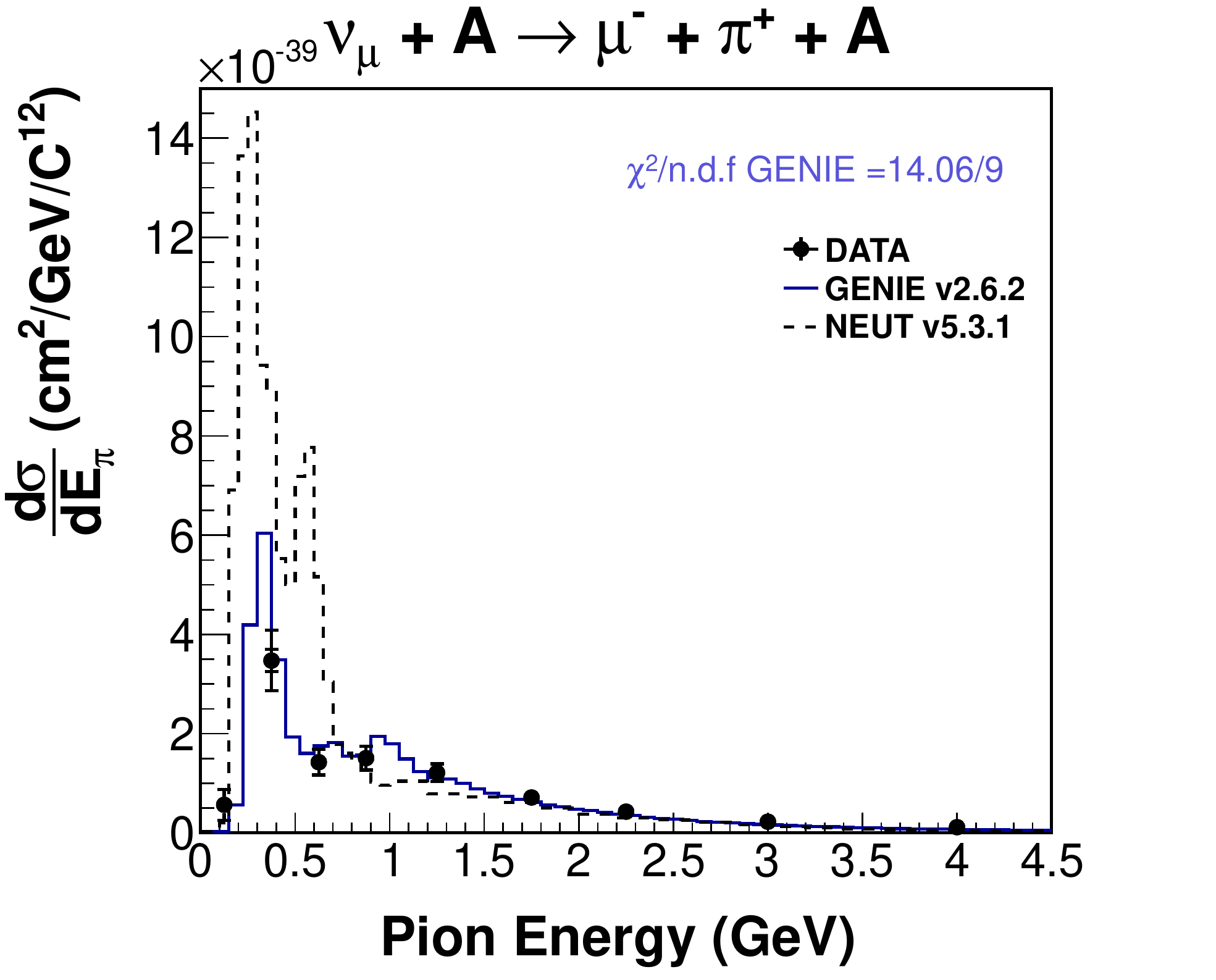}
\put (64,32) {(a)}
\end{overpic}
\begin{overpic}[trim={0 0 2.2cm 0},clip,width=\halfwidth]{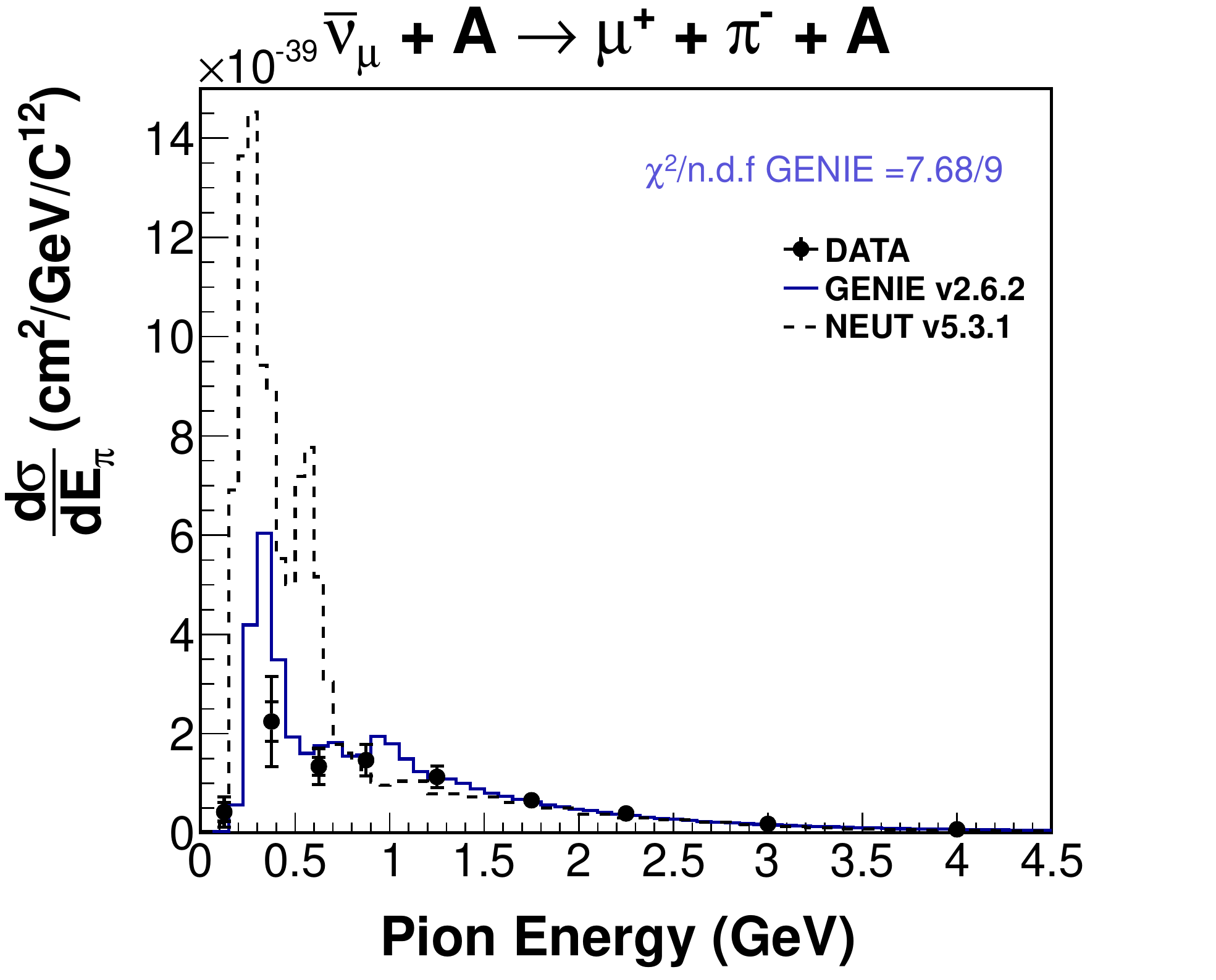}
\put (64,32) {(b)}
\end{overpic}
\caption{Charged-current coherent pion production cross sections as a function of the pion energy in (a) \numu and (b) \antinumu scattering, compared to the Rein-Sehgal model predictions implemented in \genie and \neut. Figures from Ref.~\cite{Higuera:2014azj}.}
\label{fig:cohpiEk}
\end{center}
\end{figure}

Coherent pion production is a relatively rare process that is worthy of measurement because the neutral-current channel contributes a small but poorly constrained background to electron neutrino appearance measurements. Charged-current coherent pion production, $\numuanumu+\textrm{A}\rightarrow\mu^\mp+\pi^\pm+\textrm{A}$, can be considered as the inverse process of the pion decay permitted in a nuclear environment, in analogy to pair production in QED. The momentum transfer from the pion to the nucleus---denoted by $t\equiv\left(P'-P\right)^2$ where $P'$ and $P$ are the final and initial 4-momenta of the nucleus---is small and leaves the nucleus intact, and the process has a characteristic exponential fall-off of the cross section with increasing \tabs.  By reconstructing $t$, \minerva measured this process with both \numu and \antinumu beams~\cite{Higuera:2014azj}. This measurement has provided a critical validation of the model implementation in generators (Fig.~\ref{fig:cohpiEk}). Further model comparisons in a reanalysis~\cite{Mislivec:2017qfz} show preference for the Berger-Sehgal model~\cite{Berger:2008xs} over the Rein-Sehgal model~\cite{Rein:1982pf,Rein:2006di} used
by \genie. In the low energy region, explicit nuclear model has been used to study coherent pion production (see, for example, Refs.~\cite{Singh:2006br,Alvarez-Ruso:2007irj}).

The $t$-measurement was augmented with \kplus tagging (Sect.~\ref{sec:kaon}) in order to search for the analogous kaon production, $\nu_\mu+\textrm{A}\rightarrow\mu^-+K^++\textrm{A}$. \minerva found 6 candidates with 1.77 predicted background events.  This observation comprises evidence at 3.0\,$\sigma$ that CC coherent $K^+$ production does indeed occur~\cite{Wang:2016pww}. 

A process that is the analog of neutrino NC coherent $\pi^{0}$ production on nuclei, is 
neutrino diffractive $\pi^0$ production on hydrogen: 
 $\nuanu+\textrm{H}\rightarrow\nuanu+\pi^0+\textrm{H}$. 
 This reaction relies solely on vacuum-quantum-number (Pomeron) exchange, hence diffractive~\cite{Rein:1986cd}.
 In the past, diffractive $\pi^0$ production on hydrogen was not included in neutrino event generators.  It has been identified by \minerva
 as the cause of a data excess observed in neutral-current events containing electromagnetic showers in the tracker~\cite{Wolcott:2016hws}. 

\section{Conclusions and Outlook}

With the \numi Low-Energy data, \minerva has investigated a wide variety of neutrino interactions in the GeV region of incident neutrino and antineutrino energies,
including elastic scattering on electrons as well as neutrino-nucleus 
incoherent and coherent scattering processes. The reported measurements include
utilization of neutrino scattering observations to constrain the flux, precision measurements of model parameters and model validation, and discoveries of novel processes.  The results show  neutrino-nucleus interactions to involve complex phenomena which challenge many of the current theoretical descriptions.  Incremental improvements to the models have been essential to progress with \minerva data and interpretation. The experiment's Medium-Energy data, analyses of which are currently underway, will enable
comparisons of interaction channels on a range of nuclei at new levels of statistical precision,
and expansion to kinematic phase space that has heretofore not been accessible. These data, consolidated by a preservation campaign~\cite{Fine:2020snd,Messerly:2021xpx}, will not only facilitate future precision measurements of neutrino oscillations, but also further extend 
the knowledge of electroweak phenomena that lie at the intersection of nuclear and particle physics. 

\section*{Acknowledgments}
 
This document was prepared by members of the MINERvA Collaboration using the resources of the Fermi National Accelerator Laboratory (Fermilab), a U.S. Department of Energy, Office of Science, HEP User Facility. Fermilab is managed by Fermi Research Alliance, LLC (FRA), acting under Contract No. DE-AC02-07CH11359.
These resources included support for the MINERvA construction project, and support
for construction also
was granted by the United States National Science Foundation under
Award No. PHY-0619727 and by the University of Rochester. Support for
participating scientists was provided by NSF and DOE (USA); by CAPES
and CNPq (Brazil); by CoNaCyT (Mexico); by Proyecto Basal FB 0821, CONICYT PIA ACT1413, and Fondecyt 3170845 and 11130133 (Chile); 
by CONCYTEC (Consejo Nacional de Ciencia, Tecnolog\'ia e Innovaci\'on Tecnol\'ogica), DGI-PUCP (Direcci\'on de Gesti\'on de la Investigaci\'on  - Pontificia Universidad Cat\'olica del Peru), and VRI-UNI (Vice-Rectorate for Research of National University of Engineering) (Peru); NCN Opus Grant No. 2016/21/B/ST2/01092 (Poland); by Science and Technology Facilities Council (UK); by EU Horizon 2020 Marie Skłodowska-Curie Action; by a Cottrell Postdoctoral Fellowship from the Research Corporation for Scientific Advancement; by an Imperial College London President's PhD Scholarship.  We thank the MINOS Collaboration for use of its near detector data. Finally, we thank the staff of
Fermilab for support of the beam line, the detector, and computing infrastructure.

\bibliographystyle{ieeetr}
\bibliography{main}

\end{document}